\documentclass[aps,twocolumn,showpacs]{revtex4}
\usepackage{graphicx}
\usepackage{amsmath}
\usepackage{amssymb}

\begin{document}

\title{Describing a Universal Critical Behavior in a transition from order to chaos}

\author{$^1$Edson D.\ Leonel, $^1$Mayla A. M. de Almeida, $^2$Juan Pedro Tarigo, $^2$Arturo C. Marti, $^3$Diego F. M. Oliveira}

\affiliation{$^1$Departamento de F\'isica, Unesp - Universidade Estadual Paulista - 
Av.24A. 1515, 13506-700, Rio Claro, SP, Brazil\\
$^2$Facultad de Ciencias, Universidad de la Rep\'ublica, Igua 4225, Montevideo, Uruguay\\
$^3$School of Electrical Engineering and Computer Science, University of North Dakota, Grand Forks, Avenue Stop 8357, 58202, ND, USA}

\date{\today} \widetext


\begin{abstract}
We present a comprehensive discussion of a transition from integrability to non-integrability in an oval billiard with a static boundary. This transition is controlled by a deformation parameter $\epsilon$, which modifies the boundary shape from circular, corresponding to $\epsilon=0$ and an integrable dynamics, to oval for $\epsilon\neq 0$, where non-integrability emerges. The deformation of the circular billiard gives rise to a chaotic layer that develops along a well-defined stripe in phase space. By introducing a set of transformations that isolate this chaotic stripe, we characterise the diffusive spreading of ensembles of trajectories and identify an observable, $\omega_{rms,{\rm sat}}$, which plays the role of an order parameter for the transition. For small deformations, the saturation value of the diffusion obeys the scaling law $\omega_{rms,{\rm sat}}\propto\epsilon^{\tilde{\alpha}}$, with a critical exponent $\tilde{\alpha}=0.507(2)$, vanishing continuously as $\epsilon\rightarrow 0$. The associated susceptibility, $\chi=d\omega_{rms,{\rm sat}}/d\epsilon$, diverges in the same limit, signalling the presence of critical behavior analogous to that observed in second-order (continuous) phase transitions in statistical mechanics.

\end{abstract}
\maketitle

\section{Introduction}
\label{sec1}

A billiard \cite{r1} is defined as a dynamical system composed of a single particle or an ensemble of particles moving freely inside a bounded domain and undergoing elastic collisions with its boundary \cite{r3}. The reflection law at the boundary follows from momentum conservation and corresponds to specular reflection, in which the angle formed by the particle trajectory with the tangent line at the boundary before the collision is equal to that after the collision \cite{r1}. As a consequence, the tangential component of the particle velocity is preserved at each collision, while the normal component changes sign.

Billiards with static boundaries can be classified into at least three categories \cite{r1}: (i) integrable billiards; (ii) ergodic billiards; and (iii) mixed billiards. In the first case, typical examples are the circular billiard - where the confining boundary is a circle and, in polar coordinates, $R(\theta)=R_0$ - and the elliptical billiard. Integrability in the circular billiard is associated with the conservation of mechanical energy and angular momentum: once the initial condition is specified by the incidence angle, the angular momentum remains constant along the trajectory. In the elliptical billiard, integrability follows from the conservation of mechanical energy and of the angular momenta with respect to the two foci \cite{r23}. Examples of case (ii) include the Sinai billiard \cite{sinai} - in which particles are confined to a square box of side $L$ containing a central circular scatterer of radius $R<L$ - and the Bunimovich stadium \cite{r2,r4}, whose geometry consists of two semicircles of radius $r$ connected by straight segments of length $L$. Depending on the combination of control parameters, both systems can exhibit ergodic dynamics \cite{r14}, meaning that the long-time evolution of a single trajectory explores the entire accessible phase space. In contrast, billiards of type (iii) display a mixed phase-space structure \cite{r5,r6,r7}, in which stability islands coexist with invariant spanning curves that separate distinct chaotic regions. Throughout this work, we assume that the billiard boundary can be described in polar coordinates by a radius of the form $R=R(\theta)$.

In this paper, we investigate the dynamics of an oval billiard with a static boundary whose radius in polar coordinates \cite{r23} is given by $R(\theta)=1+\epsilon\cos(p\theta)$. Our goal is to understand and characterise a particular transition observed in the dynamics as a control parameter is varied, namely the transition from integrability to non-integrability \cite{r10}. The system is governed by the deformation parameter $\epsilon$, which controls the integrable limit at $\epsilon=0$, where the phase space is completely regular \cite{r1}, and drives the system into a non-integrable regime for $\epsilon\neq 0$, leading to a mixed phase-space structure \cite{r23,r5}. In the vicinity of the transition, that is, for values of $\epsilon$ close to zero, the phase space exhibits a chaotic dynamics that develops along a well-defined region reminiscent of separatrix chaos \cite{r19}. Two initially nearby trajectories launched inside this region separate exponentially in time, yielding a positive Lyapunov exponent \cite{eckman} and thus characterising the dynamics as chaotic \cite{hilborn}.

Our aim is to understand how this chaotic component develops in phase space as the control parameter is varied. As $\epsilon$ increases, the chaotic region grows along a stripe-like structure, progressively occupying larger portions of phase space. Beyond a critical value of the control parameter, the invariant curves associated with trajectories grazing the boundary \cite{r25} at low incidence angles are destroyed \cite{r5}, allowing chaos to spread over almost the entire phase space. The manner in which chaos develops is reflected in the behavior of the dispersion of chaotic trajectories: this quantity vanishes following a power-law dependence, with an exponent between zero and one, as the system approaches the regular regime, while it saturates to a finite value for any nonzero deformation. The continuous vanishing of this observable and the divergent response to variations of the control parameter define the transition investigated here, in close analogy with phase transitions from the viewpoint of statistical mechanics \cite{r11,r12,r13}.

The investigation and characterisation of a phase transition in a dynamical system rely on the identification of suitable observables that display universal behavior near criticality \cite{r22}. At a transition, the system generally exhibits a form of symmetry breaking \cite{r17}, which may manifest either in the physical configuration or in the organisation of phase space. Identifying such a symmetry breaking provides insight into the nature of the transition and allows one to define an observable that separates regular (integrable) from irregular (non-integrable) phases. In this context, the behavior of an order parameter near the transition determines its classification, for instance as a first-order or a second-order (continuous) transition \cite{r11}. Since the present study concerns chaotic dynamics, the order parameter must be directly related to the emergence of chaos. An equally important issue is the behavior of the order parameter close to criticality \cite{r11}: whether it follows a power-law scaling and, if so, the value of the associated critical exponent. Knowledge of the critical exponent makes it possible to identify whether different systems exhibit the same behavior near criticality, thus defining a universality class \cite{barabasi}. Furthermore, understanding which control parameter governs the onset of chaotic diffusion allows one to identify the elementary excitation \cite{r17} responsible for transport, as well as the effective step length of the diffusion process. Finally, it is essential to determine how diffusion unfolds in phase space, whether it is uniformly distributed or trapped near specific regions \cite{r5,r16}. This information reveals the role of topological defects \cite{r17} that influence particle transport, affect ergodicity, and shape the global dynamics of the system.

Having established the general framework, we now outline the sequence of steps adopted to address the problem. 
(1) We first identify and describe the symmetry of the system that is destroyed at the transition and analyse how this symmetry breaking affects the organisation of phase space and the dynamical evolution. 
(2) We then describe how chaos develops in phase space and how chaotic diffusion progressively fills the configurational space. This characterisation allows us to determine an appropriate order parameter, analyse its behavior as the control parameter is varied, and examine the evolution of its response to changes in the control parameter, which defines the corresponding susceptibility. 
(3) As the investigation proceeds, a key point is the identification of the elementary excitation of the system that drives particles to move chaotically through phase space. 
(4) Finally, we investigate whether specific regions of phase space significantly influence the particle dynamics and thereby modify the associated probability distribution function. In particular, we address the question of whether topological defects exist in phase space and how they alter the ballistic motion of particles.

The sequence of steps outlined above is organised in sections throughout the paper. In Sec.~\ref{sec2}, we define the model and describe how the system is evolved from different initial conditions. The discussion of symmetry breaking and its impact on the dynamics is presented in Sec.~\ref{sec3}. The identification of the order parameter, its scaling behavior, and the influence of the control parameter on the susceptibility are addressed in Sec.~\ref{sec4}. The analysis of the elementary excitation responsible for chaotic diffusion is the subject of Sec.~\ref{sec5}, while the role of topological defects - which leads naturally to a discussion of stickiness in phase-space diffusion - is examined in Sec.~\ref{sec6}. 

We then move beyond the specific model in Section \ref{sec7} and discuss the universality of the transition by making a direct comparison between the results obtained here and those reported for the Fermi-Ulam model \cite{r8}, for periodically corrugated waveguides \cite{r9,bliock}, and for a family of two-dimensional area-preserving maps \cite{r10} formulated in terms of angle-action variables, in which the angle diverges in the limit of vanishing action \cite{farhan}. In this context, we also discuss saturation effects for the diffusion and interpret them in terms of an effective correlation length associated with the transition. Finally, in Sec.~\ref{sec8}, we present a broader discussion and outline a general framework for extending the present investigation to other models sharing similar dynamical features and a perspective for future research. Our conclusions are made in Section \ref{sec9}.

\section{The model and the mapping}
\label{sec2}

The model considered in this work is the oval billiard \cite{r23,r5}, whose boundary in polar coordinates is defined by
$R(\theta)=1+\epsilon\cos(p\theta)$, where $\theta$ is the polar angle, $\epsilon$ controls the deformation with respect to the circular shape, and $p$ is an integer parameter.

\begin{figure}[t]
\centerline{\includegraphics[width=0.8\linewidth]{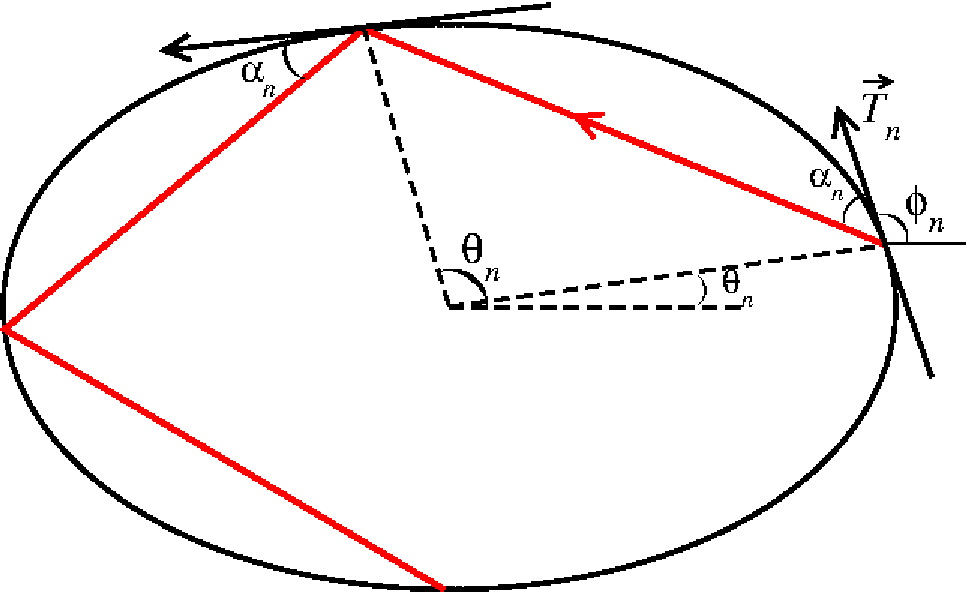}}
\caption{Sketch of the angles describing the billiard dynamics. The boundary is constructed using $R=1+\epsilon\cos(p\theta)$ with $p=2$ (assumed constant throughout this work) and $\epsilon=0.05$. The variable $\theta_n$ denotes the polar angle at the $n$th collision, $\vec{T}_n$ is the tangent vector at $\theta_n$, and $\alpha_n$ is the angle of the trajectory measured with respect to the tangent vector $\vec{T}_n$.}
\label{Fig1}
\end{figure}

The dynamics of a particle confined within this boundary is described by a discrete mapping in the variables $(\theta_n,\alpha_n)$, where $\theta_n$ denotes the angular position of the particle at the $n$th collision with the boundary and $\alpha_n$ is the angle between the particle trajectory and the tangent vector to the boundary at $\theta_n$, as illustrated in Fig.~\ref{Fig1}. The subscript $n$ labels successive collisions of the particle with the boundary.

Since the boundary is expressed in polar coordinates, the position of the particle at the $n$th collision, written in rectangular coordinates, is given by
$X(\theta_n)=R(\theta_n)\cos(\theta_n)$ and 
$Y(\theta_n)=R(\theta_n)\sin(\theta_n)$. For a given initial condition $(\theta_n,\alpha_n)$, the angle between the tangent vector to the boundary at the point $(X(\theta_n),Y(\theta_n))$ and the horizontal axis is
$\phi_n=\arctan\!\left[Y'(\theta_n)/X'(\theta_n)\right]$. Between two successive collisions with the boundary, no forces act on the particle, so it moves along a straight line with constant velocity. The equation describing the trajectory between collisions is therefore
\begin{eqnarray}
Y(\theta_{n+1}) - Y(\theta_n) = \tan(\alpha_n + \phi_n)\left[
X(\theta_{n+1}) - X(\theta_n) \right],
\label{B_eq2}
\end{eqnarray}
where $\phi_n$ is the slope of the tangent vector measured with respect to the positive $X$ axis, and $X(\theta_{n+1})$ and $Y(\theta_{n+1})$ are the new rectangular coordinates obtained by solving Eq.~(\ref{B_eq2}). The angle between the trajectory and the tangent vector at the subsequent collision point $\theta_{n+1}$ is then given by
\begin{eqnarray}
\alpha_{n+1} = \phi_{n+1} - (\alpha_n + \phi_n).
\label{B_eq3}
\end{eqnarray}

The mapping that governs the particle dynamics is defined by the iteration of the operator $\bf{T}$,
\begin{equation}
\bf{T}:\left\{
\begin{array}{ll}
H(\theta_{n+1}) = R(\theta_{n+1})\sin(\theta_{n+1}) - Y(\theta_n) - \\
~~~~~~\tan(\alpha_n + \phi_n)\left[ R(\theta_{n+1})\cos(\theta_{n+1}) 
- X(\theta_n) \right], \\
\alpha_{n+1} = \phi_{n+1} - (\alpha_n + \phi_n),
\end{array}
\right.
\label{B_eq4}
\end{equation}
where $\theta_{n+1}$ is obtained numerically from the condition $H(\theta_{n+1})=0$ and
$\phi_{n+1}=\arctan\!\left[Y'(\theta_{n+1})/X'(\theta_{n+1})\right]$.

Starting from an initial condition $(\theta_0,\alpha_0)$, a single application of the operator $\bf{T}$ yields ${\bf{T}}(\theta_0,\alpha_0)=(\theta_1,\alpha_1)$. Subsequent iterations generate the sequence $(\theta_2,\alpha_2)$, $(\theta_3,\alpha_3)$, and so on. Symbolically, this evolution can be represented as
$n \xrightarrow{\bf{T}} (n+1)\xrightarrow{\bf{T}} (n+2)\xrightarrow{\bf{T}} (n+3)\ldots$,
which defines an orbit originating from the initial condition $(\theta_0,\alpha_0)$. A large ensemble of initial conditions allows one to construct the phase space (or configurational space) and to characterise the different dynamical regimes exhibited by the system.

The numerical procedure adopted to solve the equation $H(\theta_{n+1})=0$ and thus obtain $\theta_{n+1}$ is described as follows \cite{num1}. Starting from an initial condition $(\theta_0,\alpha_0)$, the position of the particle on the boundary and the slope of its trajectory are known. We then divide the interval $\theta\in[0,2\pi]$ into $m$ equal subintervals, defining an increment $\delta\theta=2\pi/m$, and apply the bisection method \cite{num2} to locate the roots of $H(\theta)$. Specifically, we search for intervals satisfying the condition
$H(i\delta\theta)H[(i+1)\delta\theta]<0$, with $i=0,\ldots,m-1$. When this condition is fulfilled, it indicates that the interval $\theta\in[i\delta\theta,(i+1)\delta\theta]$ contains a root. The interval is then iteratively reduced by halving its size, as prescribed by the bisection method \cite{num1,num2}, until a predefined numerical accuracy of $10^{-12}$ is achieved. Other numerical accuracies were also considered with no change in any result.

\begin{figure}[t]
\centerline{\includegraphics[width=1.0\linewidth]{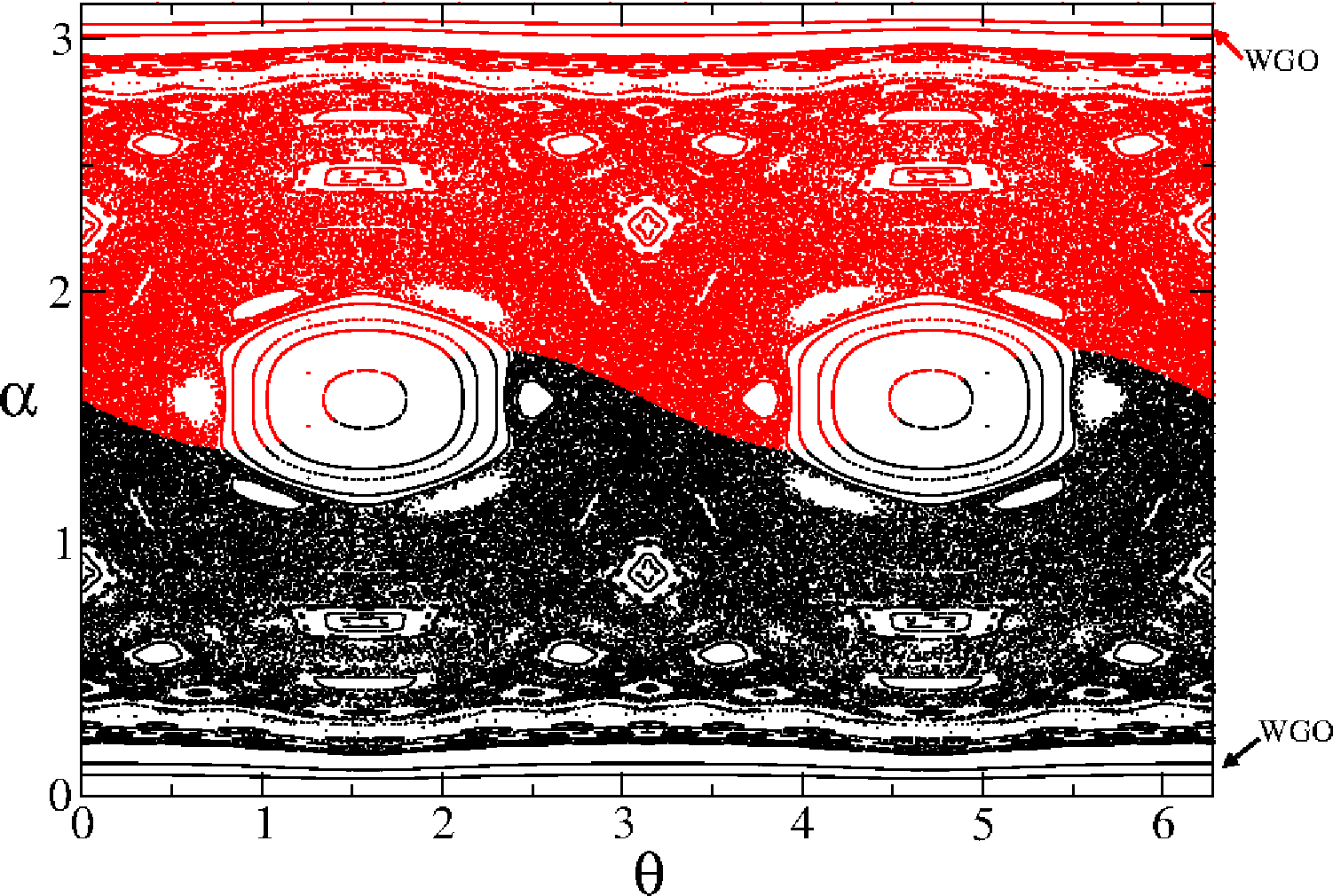}}
\caption{(Color online) Phase-space of the mapping~(\ref{B_eq4}) for the control parameter $\epsilon=0.1$. A mixed phase-space structure is observed, with periodic islands surrounded by a chaotic sea bounded by a set of whispering gallery orbits (WGO), as indicated in the figure.}
\label{Fig2}
\end{figure}

Once a solution is obtained, the procedure is repeated over the remaining intervals in order to locate a second root. For the parameter range considered here, $\epsilon<\epsilon_c$, there are always two solutions: one corresponding to the point from which the particle departed and another corresponding to the next collision point. The former is discarded, and the latter is retained to update the mapping. To motivate the discussion of the critical parameter $\epsilon_c$, we first present a phase-space for the control parameter $\epsilon=0.1$, shown in Fig.~\ref{Fig2}.

The phase space obtained for $\epsilon=0.1$ exhibits three distinct dynamical components: (i) periodic regions associated with stability islands; (ii) a set of trajectories moving close to the boundary with low incidence angles, known as whispering gallery orbits (WGO) \cite{r25}, as indicated in the figure; and (iii) a chaotic domain, represented by a cloud of points. The colour scheme distinguishes trajectories with positive angular momentum, which move counter-clockwise (black), from those with negative angular momentum, which move clockwise (red). The angular momentum is computed as $\vec{\ell}=\vec{r}\times\vec{p}$, where $\vec{r}$ denotes the particle position and $\vec{p}$ its momentum, yielding $\vec{\ell}= \hat{k}(x\dot{y}-y\dot{x})$ and $\hat{k}$ denotes the unity vector perpendicular to the billiard plane.

The chaotic sea shown in Fig.~\ref{Fig2} is generated by evolving a single initial condition over a long time. To confirm the chaotic nature of this region, we compute the largest Lyapunov exponent \cite{eckman} for five different initial conditions chosen within the chaotic domain, obtaining $\lambda = 0.25(2)$. A detailed description of the algorithm used to calculate the Lyapunov exponent is provided in Appendix~\ref{Aapp1}.

The invariant curves near the boundary, corresponding to trajectories with low incidence angles, persist up to a critical value of the control parameter, namelly $\epsilon_c$. Their observation was made very earlier than the billiard investigation, far more than a century ago by Rayleigh \cite{r25}, as reflecting waves in catedrals. Their existence is related to a geometric property of the phase space and ceases when the curvature of the boundary changes sign. The curvature is given by
\begin{equation}
\kappa(\theta)=\frac{X'(\theta)Y''(\theta)-X''(\theta)Y'(\theta)}
{\left[X'^2(\theta)+Y'^2(\theta)\right]^{3/2}},
\end{equation}
where primes denote derivatives with respect to $\theta$. The condition $\kappa(\theta)=0$ yields a critical value of the control parameter \cite{r5},
\begin{equation}
\epsilon_c=\frac{1}{1+p^2}.
\end{equation}
For $\epsilon>\epsilon_c$, the invariant spanning curves associated with low-incidence-angle trajectories, which generate the whispering gallery orbits, are destroyed.

\section{Symmetry breaking}
\label{sec3}

\begin{figure}[t]
\centerline{(a)\includegraphics[width=0.8\linewidth]{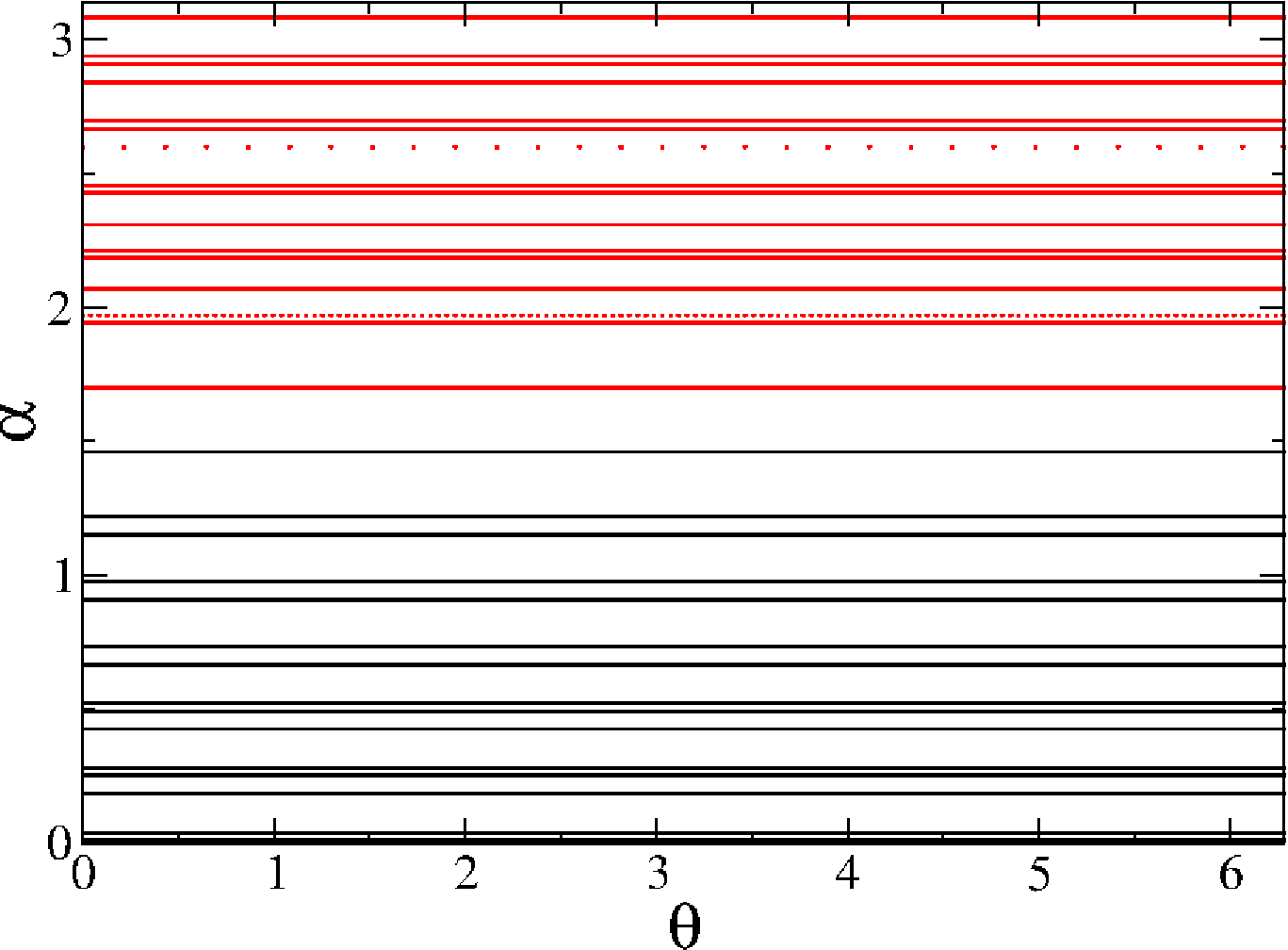}}
\centerline{(b)\includegraphics[width=0.8\linewidth]{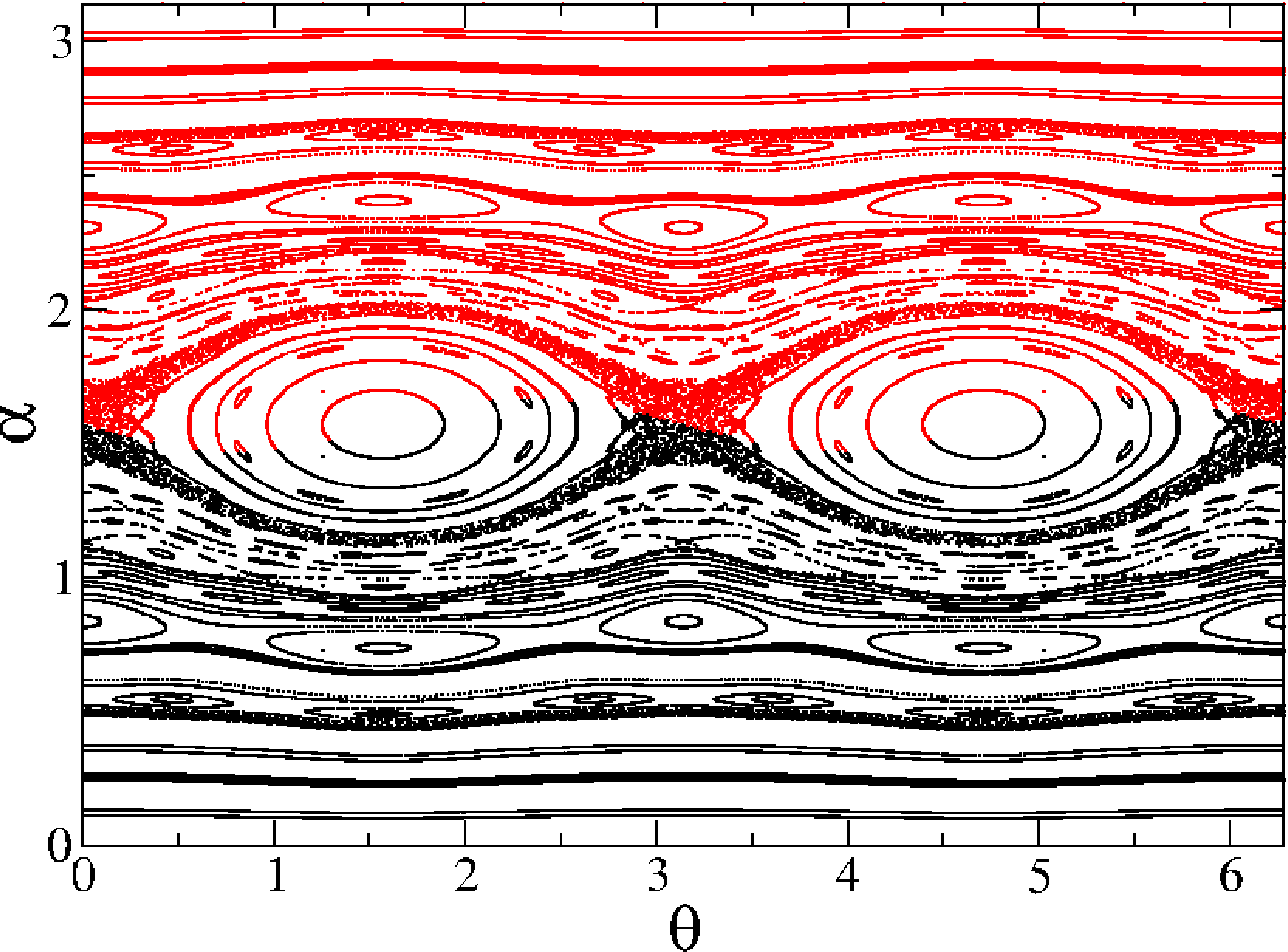}}
\caption{(Color online) Plot of the phase-space for mapping~(\ref{B_eq4}) for the control parameter: (a) $\epsilon=0$, where a set of regular orbits is observed; (b) $\epsilon=0.05$, where a mixed phase-space structure emerges, consisting of periodic islands surrounded by a chaotic sea bounded by whispering gallery orbits.}
\label{Fig3}
\end{figure}

Having established the dynamical description of the billiard in terms of a discrete map, we now discuss the structure of phase space and identify the symmetry that is destroyed at the transition. Figure~\ref{Fig3}(a) shows the phase-space portrait for the control parameter $\epsilon=0$. In this case, only periodic dynamics, represented by discrete sets of points, and quasiperiodic dynamics, represented by continuous invariant curves, are observed.

The same colour scheme is adopted throughout. Figure~\ref{Fig3}(a) clearly shows that the phase space is entirely filled by periodic or quasiperiodic trajectories. Once a particle starts moving clockwise, it continues to do so indefinitely, and the same holds for counter-clockwise motion. Together with the conservation of mechanical energy, the conservation of angular momentum implies the existence of two independent constants of motion, rendering the system integrable \cite{r24}.

This situation changes as soon as the control parameter departs from zero. For $\epsilon\neq 0$, the regular phase-space structure observed in Fig.~\ref{Fig3}(a) is destroyed and replaced by a mixed structure, as shown in Fig.~\ref{Fig3}(b). In this regime, regular dynamics persists in the form of whispering gallery orbits and stability islands, while chaotic motion develops in extended regions of phase space. In particular, near $\alpha=\pi/2$, a chaotic layer emerges with a structure reminiscent of separatrix chaos \cite{r19}. Depending on the region of phase space, both chaotic trajectories and regular island dynamics may exhibit changes in the sign of the angular momentum, indicating that orbits can reverse their direction of motion from counter-clockwise to clockwise, or vice versa, during the evolution.

Since the angular momentum is no longer conserved for $\epsilon\neq 0$, integrability is lost. The symmetry present in Fig.~\ref{Fig3}(a), manifested by the invariance of trajectories at constant $\alpha$, is absent in Fig.~\ref{Fig3}(b), where the mixed phase-space structure dominates. The transition from a fully regular phase space to a mixed one therefore corresponds to a symmetry breaking in phase space, which identifies both the symmetry lost at the transition and the dynamical invariant that ceases to be conserved.

\section{Order parameter and its susceptibility}
\label{sec4}

\begin{figure}[t]
\centerline{(a)\includegraphics[width=0.475\linewidth]{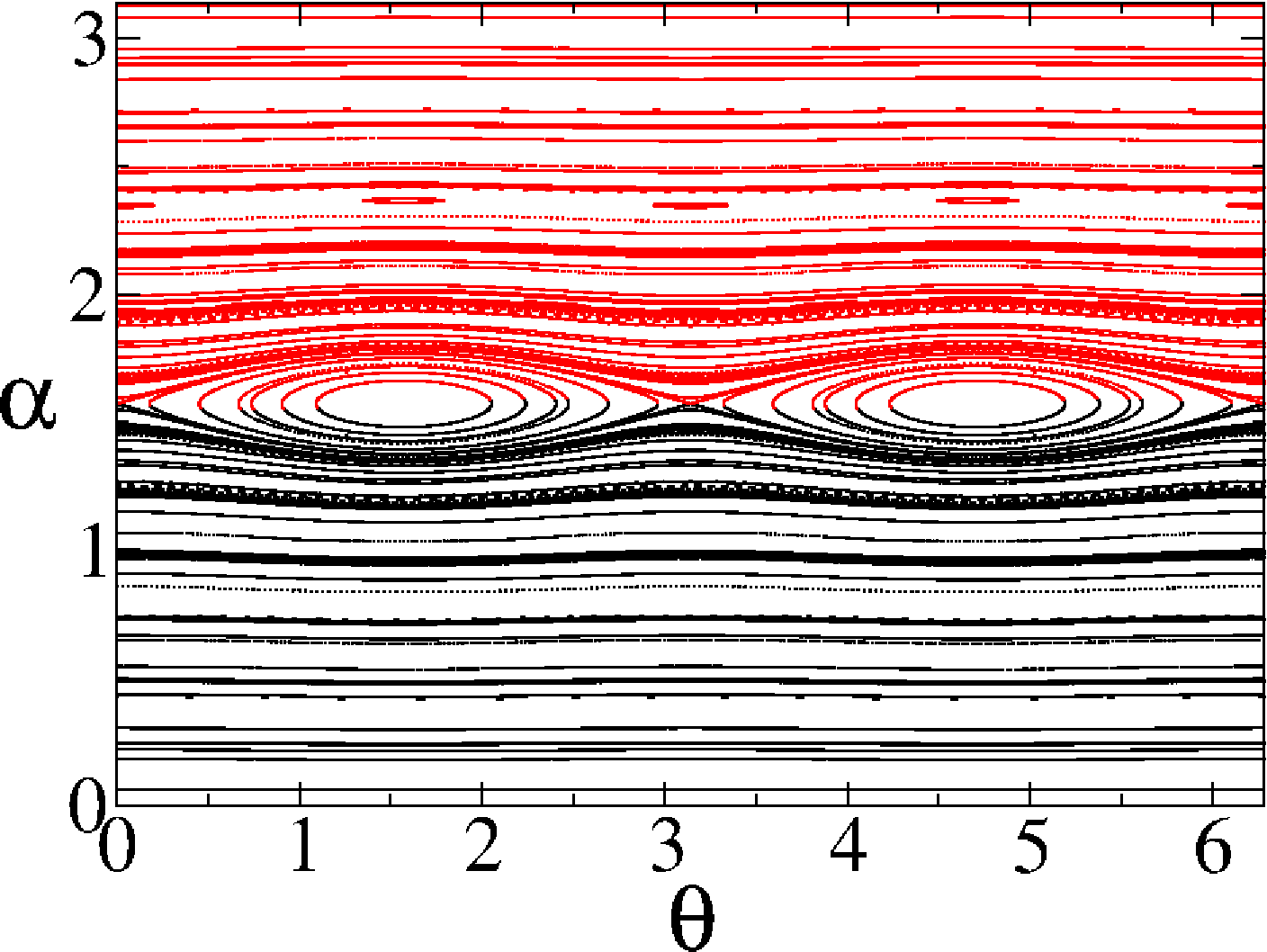}
(b)\includegraphics[width=0.475\linewidth]{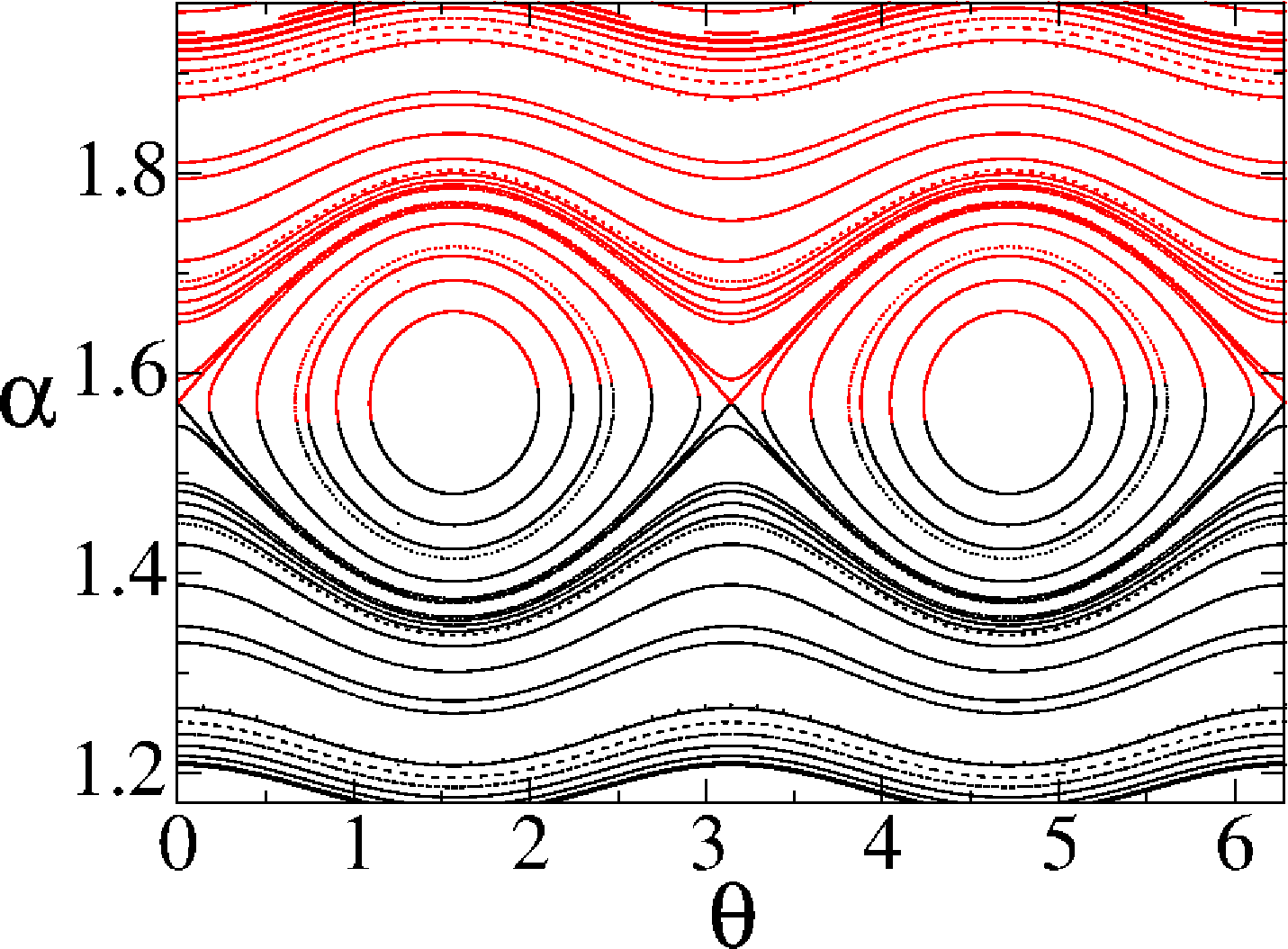}}
\centerline{(c)\includegraphics[width=0.475\linewidth]{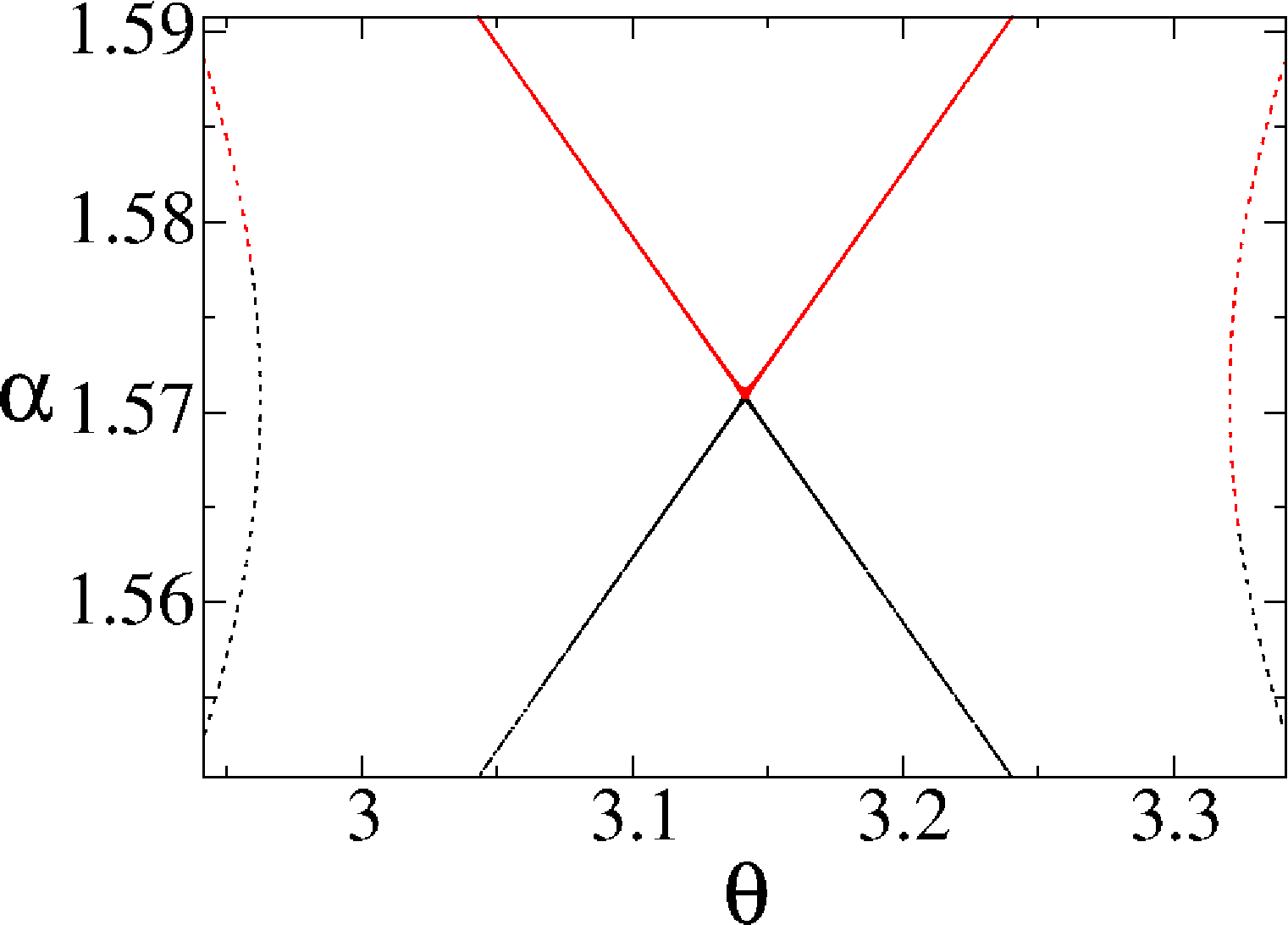}
(d)\includegraphics[width=0.475\linewidth]{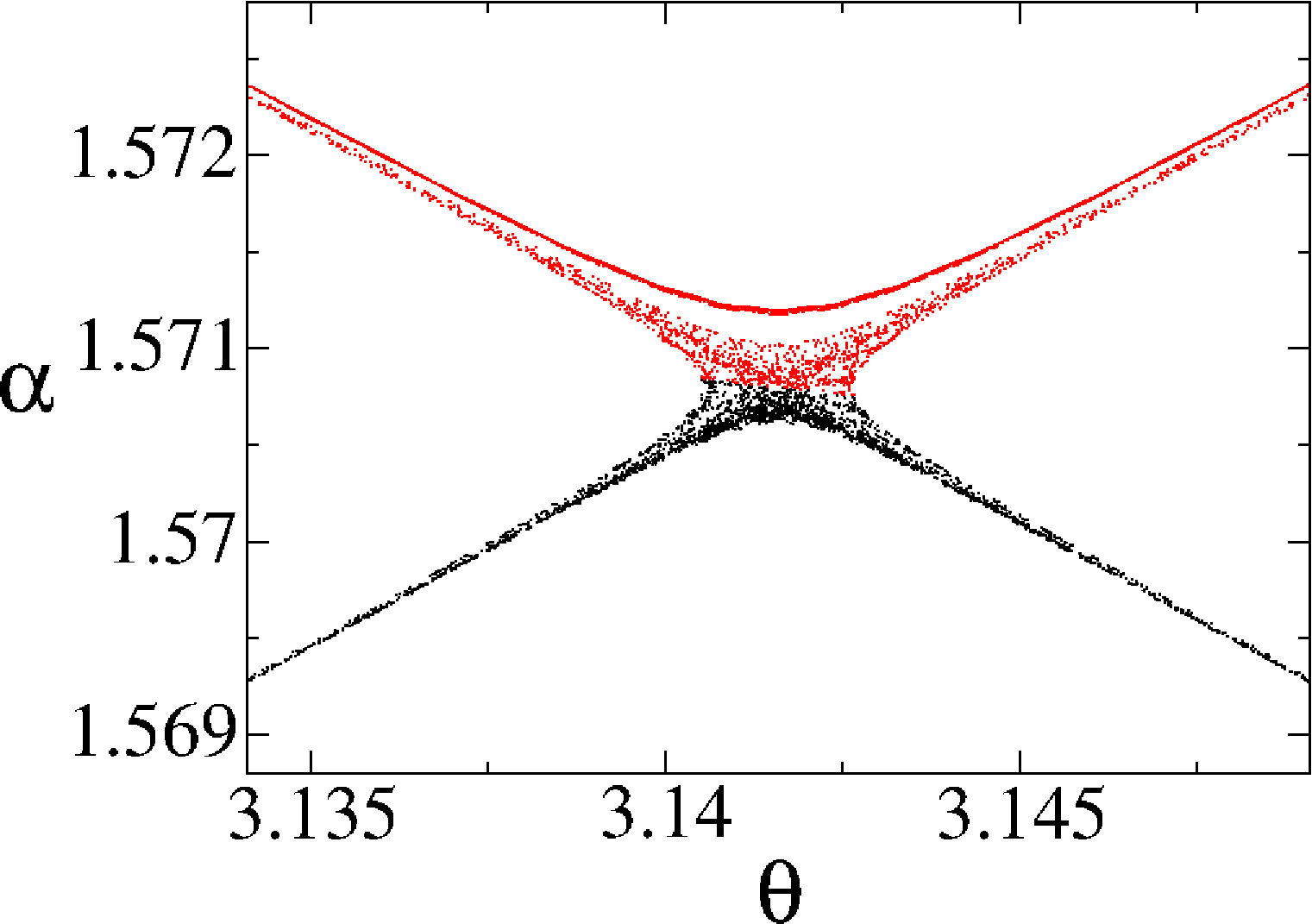}}
\caption{Phase-space portraits of the mapping~(\ref{B_eq4}) for the control parameter $\epsilon=0.01$ at different levels of magnification: (a) $\theta\in[0,2\pi]$ and $\alpha\in[0,\pi]$, showing the full phase space; (b) $\theta\in[0,2\pi]$ and $\alpha\in[\pi/2-0.4,\pi/2+0.4]$, highlighting the region near the islands and revealing a separatrix-like structure; (c) $\theta\in[\pi-0.2,\pi+0.2]$ and $\alpha\in[\pi/2-0.02,\pi/2+0.02]$, where the separatrix-like structure is seen to occupy a finite area; (d) $\theta\in[\pi-0.0075,\pi+0.0075]$ and $\alpha\in[\pi/2-0.002,\pi/2+0.002]$, showing a chaotic region forming a narrow chaotic stripe near the saddle point $(\theta,\alpha)=(\pi,\pi/2)$.}
\label{Fig4}
\end{figure}

In this section, we identify the observable that plays the role of an order parameter \cite{r11} for the transition. Such an observable must clearly distinguish the fully regular regime from the regime in which chaotic dynamics is present. For the oval billiard considered here, chaos initially emerges in a structure reminiscent of separatrix chaos. To illustrate how this type of chaotic dynamics develops, Fig.~\ref{Fig4}(a) shows the phase-space portrait for $\epsilon=0.01$. Since the control parameter is relatively small, the phase space appears, at first glance, predominantly regular, consisting of a set of invariant curves associated with whispering gallery orbits and two large stability islands located near $(\theta,\alpha)=(\pi/2,\pi/2)$ and $(\theta,\alpha)=(3\pi/2,\pi/2)$. In the central region of the phase space, a structure resembling a separatrix curve can be identified at this scale. However, this apparent separatrix is not a true invariant curve.

A closer inspection reveals a more complex scenario. Figure~\ref{Fig4}(b) shows a magnified view of Fig.~\ref{Fig4}(a) near the island region. The separatrix-like structure becomes more apparent, coexisting with invariant curves forming whispering gallery orbits and those surrounding the islands. A further zoom, shown in Fig.~\ref{Fig4}(c), demonstrates that this structure is not a line but instead occupies a finite area of phase space. Finally, an even deeper magnification, displayed in Fig.~\ref{Fig4}(d), clearly reveals that this region is chaotic, forming a narrow chaotic stripe in the phase-space portrait.

As the control parameter $\epsilon$ increases, the chaotic domain develops into a well-defined layer that thickens progressively, forming an extended chaotic stripe in phase space. In Fig.~\ref{Fig4}(d), corresponding to $\epsilon=0.01$, the chaotic layer is very thin, whereas it becomes substantially broader for $\epsilon=0.05$, as shown in Fig.~\ref{Fig3}(b), and eventually dominates almost the entire phase space for $\epsilon=0.1$, as illustrated in Fig.~\ref{Fig2}. As the chaotic layer grows, increasingly large regions of phase space become chaotic. The transition from integrability at $\epsilon=0$ to non-integrability for any $\epsilon\neq 0$ can therefore be investigated by analysing the dynamics of particles evolving within this chaotic domain. This approach allows us to characterise the diffusion of particles in phase space and to study the transition from integrability to non-integrability in the oval billiard.

\begin{figure}[t]
\centerline{\includegraphics[width=1.0\linewidth]{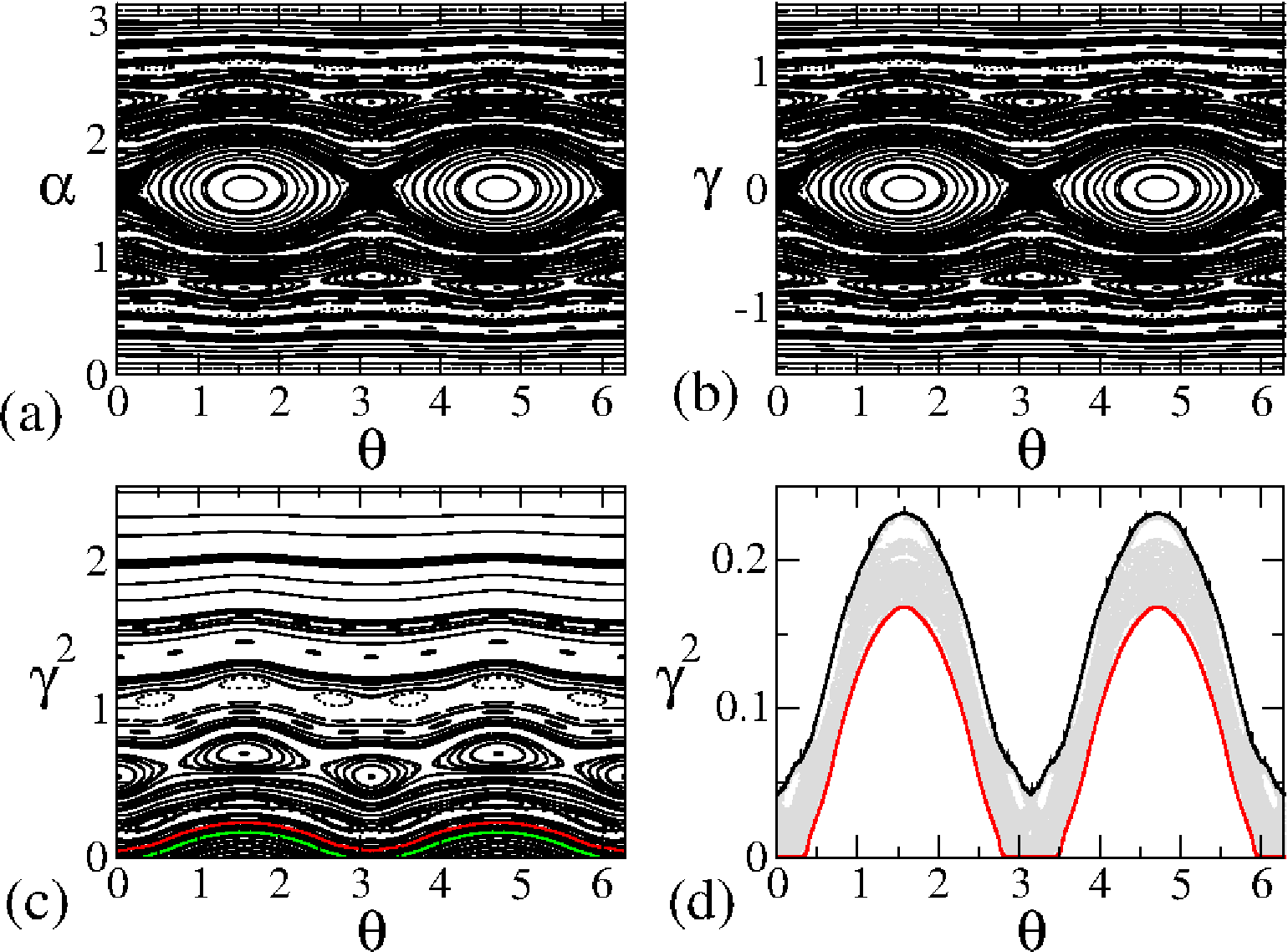}}
\caption{(a) Phase-space portrait for $\epsilon=0.05$, showing a chaotic layer symmetric with respect to $\alpha=\pi/2$. (b) Same as (a) after the transformation $\gamma=\alpha-\pi/2$, revealing symmetry with respect to $\gamma=0$. (c) Same as (b) after the transformation $\gamma\rightarrow \gamma^2$, which folds the negative part of the phase space onto the positive one. (d) Envelopes of the stripe delimiting the chaotic layer in phase space.}
\label{Fig5}
\end{figure}

We now turn to the description of diffusion along the chaotic layer. Figure~\ref{Fig5}(a) shows the phase-space portrait of the oval billiard for $\epsilon=0.05$, where a well-developed chaotic layer is clearly visible and exhibits symmetry with respect to $\alpha=\pi/2$. Figure~\ref{Fig5}(b) shows the same phase space after the transformation $\gamma=\alpha-\pi/2$, which makes the symmetry of the chaotic layer explicit with respect to the origin $\gamma=0$.

Since we are interested in average properties of the diffusion process and the chaotic layer is symmetric with respect to zero, the negative part of phase space is statistically equivalent to the positive one. As a consequence, the time average of any observable that is odd in $\gamma$ converges to zero for sufficiently long times, in particular the average angle. To avoid this cancellation and to focus on the amplitude of the chaotic diffusion, we transform the phase space shown in Fig.~\ref{Fig5}(b) by squaring the variable, $\gamma\rightarrow\gamma^2$, leading to Fig.~\ref{Fig5}(c), where the chaotic layer is mapped entirely onto the positive part of phase space. Figure~\ref{Fig5}(d) shows the envelope of the chaotic layer, which delimits the region where chaotic dynamics takes place.

\begin{figure}[t]
\centerline{\includegraphics[width=0.6\linewidth]{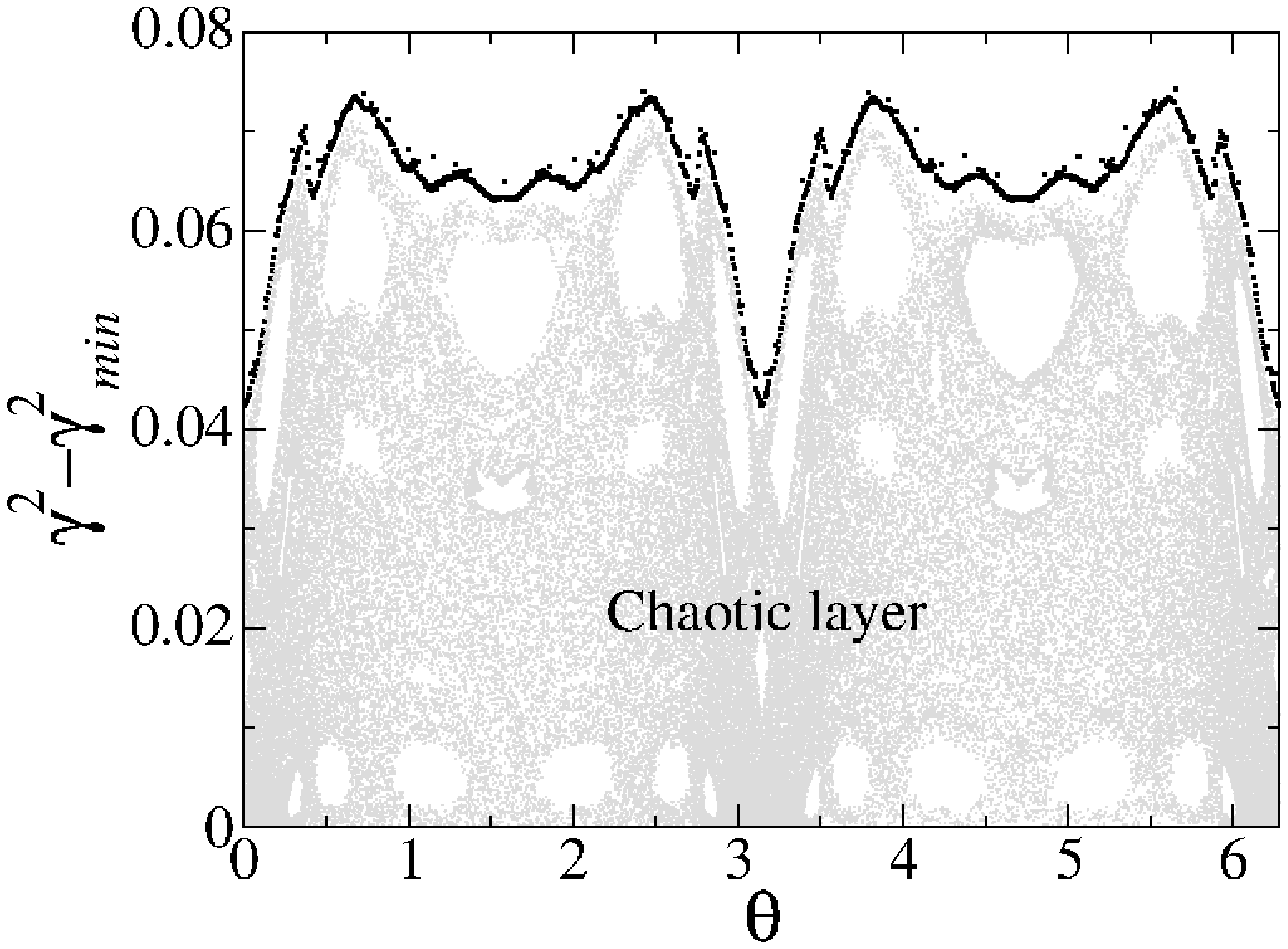}}
\caption{Color online) Plot of the chaotic layer after the transformation $\gamma^2-\gamma_{min}^2$. The bullet points correspond to a numerical approximation of the chaotic border. The control parameters are $\epsilon=0.05$.}
\label{stripe}
\end{figure}

Since we are interested in measuring the deviation along a chaotic stripe, a further transformation, $\omega=\gamma^2-\gamma_{min}^2$, is particularly useful. It allow for analysing the width of the chaotic diffusion, as illustrated in Fig.~\ref{stripe}.

\begin{figure}[t]
\centerline{(a)\includegraphics[width=0.8\linewidth]{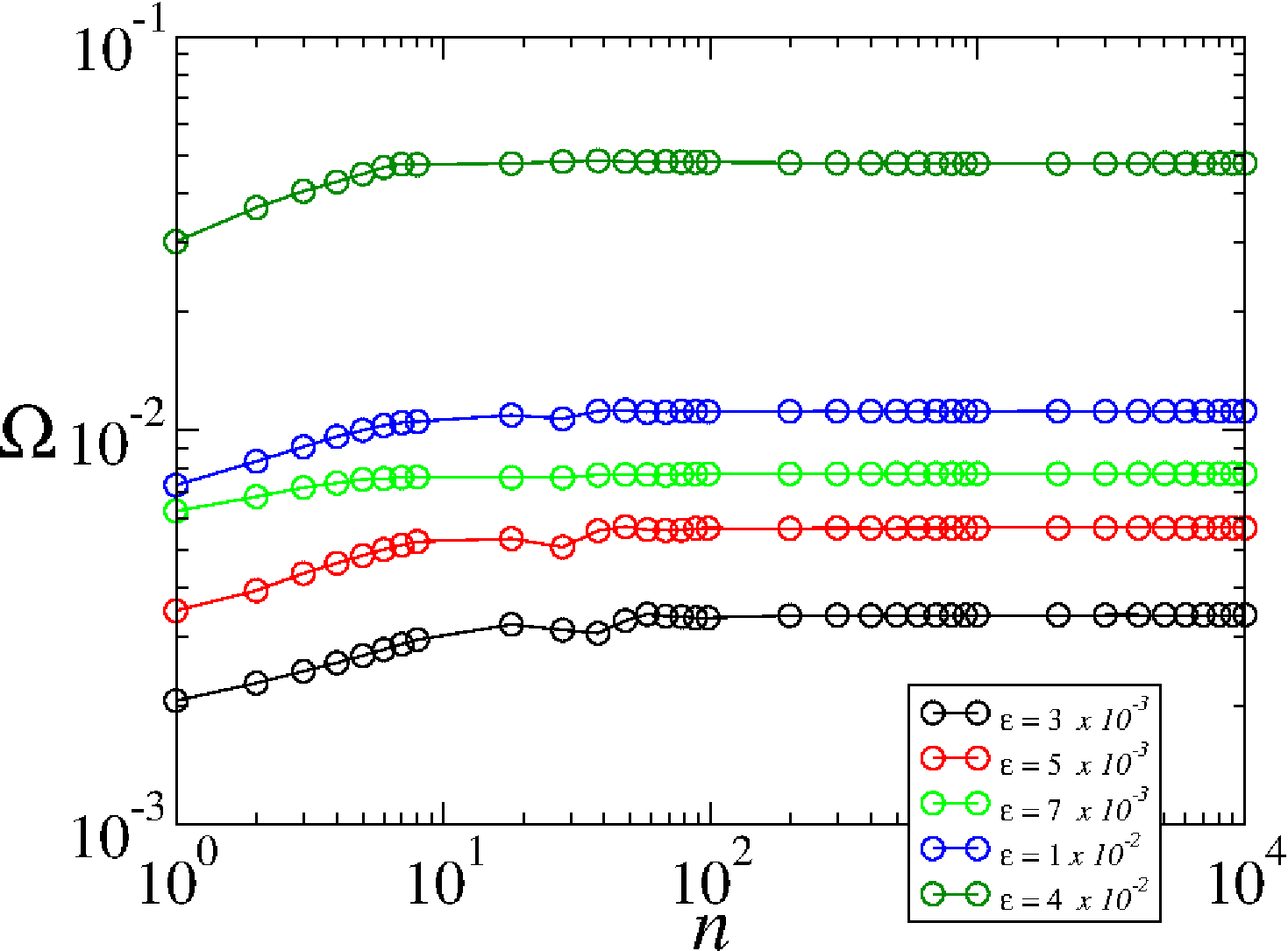}}
\centerline{(b)\includegraphics[width=0.75\linewidth]{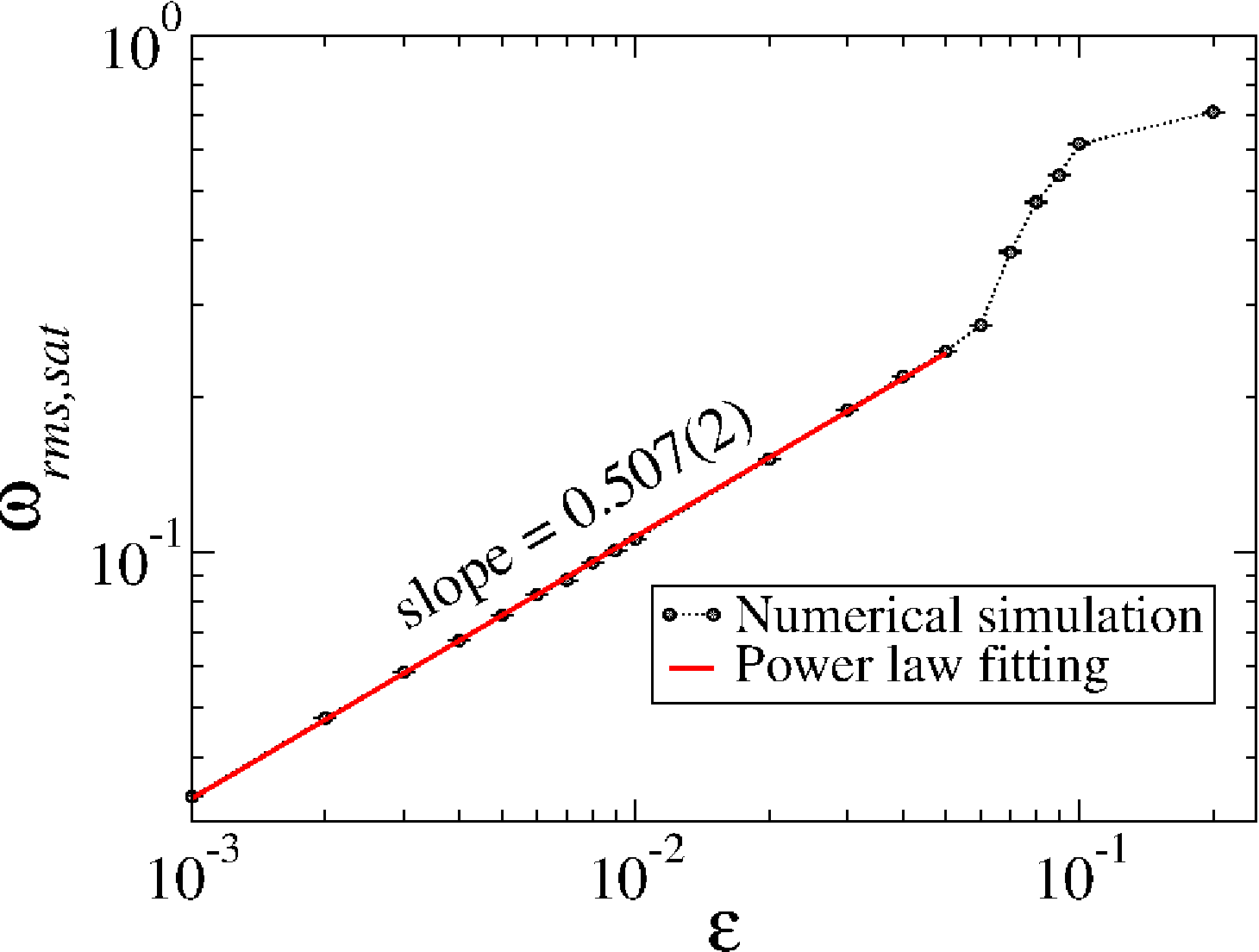}}
\caption{(Color online) (a) Plot of $\Omega$ as a function of the number of collisions $n$ for different values of the control parameter $\epsilon$, as labelled in the figure. (b) Plot of the saturation value $\omega_{rms,{\rm sat}}$ as a function of $\epsilon$. A power-law fit yields the exponent $\tilde{\alpha}=0.507(2)$. Error bars corresponding to the statistical uncertainty of the ensemble averages are also shown.}
\label{Fig6}
\end{figure}

We now investigate the behavior of the deviation around the chaotic stripe as a function of the control parameter $\epsilon$. This quantity is defined as
\begin{equation}
\Omega(n,\epsilon)=\frac{1}{M}\sum_{i=1}^{M}
\sqrt{\overline{\omega_i^2}(n,\epsilon)-\overline{\omega_i}^2(n,\epsilon)},
\end{equation}
where $\overline{\omega}(n,\epsilon)=\frac{1}{n}\sum_{k=1}^{n}\omega_k$. We consider an ensemble of $M=10^3$ different initial conditions, with initial angles $\alpha$ chosen very close to the lower boundary of the chaotic stripe and $\theta_0$ uniformly distributed in the interval $\theta_0\in[0,2\pi]$. This choice allows particles to experience maximal diffusion within the chaotic region. Figure~\ref{Fig6}(a) shows a plot of $\Omega$ as a function of the number of collisions $n$ for different values of the control parameter $\epsilon$, as indicated in the figure. We observe that the deviation initially grows with $n$ and, after a crossover, eventually saturates, signalling the onset of a steady state of the diffusive process.

Since $\Omega$ is defined in terms of the squared angular variable, it is convenient to analyse instead the quantity $\omega_{rms}=\sqrt{\Omega}$, which has the same units as the angle. For sufficiently large values of $n$, all curves for $\omega_{rms}$ approach a saturation regime. A direct measurement of the saturation values for different values of $\epsilon$ is presented in Fig.~\ref{Fig6}(b). For small $\epsilon$, a clear power-law behavior is observed, except for values $\epsilon\gtrsim0.07$, where the chaotic domain rapidly expands in phase space and dominates the dynamics. This latter regime lies far from the vicinity of integrability, which is the focus of the present critical analysis.

A power-law fit to the data points closest to criticality yields
$\omega_{rms,{\rm sat}}\propto \epsilon^{\tilde{\alpha}}$ with $\tilde{\alpha}=0.507(2)$. Since we are analysing a diffusive process, it is natural for the order parameter to quantify the extent of diffusion \cite{r10} along the chaotic stripe. The observable $\omega_{rms,{\rm sat}}$ satisfies all the necessary criteria to play this role: it is well defined within the chaotic layer, follows a power-law dependence on $\epsilon$ near the transition, and vanishes continuously and monotonically as $\epsilon\rightarrow 0$. 

The response of the order parameter to variations in the control parameter, namely the susceptibility, is given by
$\chi=\frac{d\omega_{rms,{\rm sat}}}{d\epsilon}\propto \tilde{\alpha}\,\epsilon^{\,\tilde{\alpha}-1}$. Since $0<\tilde{\alpha}<1$, the susceptibility diverges in the limit $\epsilon\rightarrow 0$. These results demonstrate that the transition from integrability to non-integrability in the oval-like billiard displays the characteristic features of a second-order (continuous) phase transition.

\section{Elementary excitation}
\label{sec5}

We now turn to the identification of the elementary excitation underlying the dynamics. For $\epsilon=0$, no diffusion is observed in phase space, which remains entirely regular \cite{r1}. When $\epsilon\neq 0$, however, a chaotic layer emerges and propagates along a stripe in phase space. This behavior occurs exclusively because the control parameter $\epsilon$ is nonzero. We therefore conclude that $\epsilon$ plays the role of an elementary excitation for the dynamics, setting the effective step length that allows diffusive motion to develop in phase space.

Since the diffusion is measured in the dispersion of the angle $\alpha$ and by virtue of the fact it is obtained directly from the expression of $\phi=\arctan\!\left[Y'(\theta)/X'(\theta)\right]$, when we substitute the terms $Y'(\theta)$ and $X'(\theta)$, proceed with a Taylor expansion near $\epsilon\approx 0$ and consider only first correction, we obtain
\begin{equation}
\phi=\arctan\left(-{{\cos(\theta)}\over{\sin(\theta)}}\right)+
\epsilon p{{\sin(p\theta)}\over{\sin(\theta)^2}}D_z\arctan(z),
\end{equation}
where $z=-{{\cos(\theta)}\over{\sin(\theta)}}$ and $D_z$ corresponds to the derivative of the $\arctan{z}$. The first correction of the angle $\phi$ is dependent on $\epsilon$, therefore confirming it as the term responsible for the diffusion of chaotic orbits in the phase space.

In statistical mechanics \cite{r11,r12,r13}, the choice of the appropriate thermodynamic potential is dictated by the constraints imposed on the system by its environment \cite{r14}, rather than by which quantities happen to fluctuate. Different statistical ensembles naturally arise depending on whether the system exchanges energy, particles, or volume with an external reservoir \cite{edlgeneris}.

For an isolated system with fixed energy $E$, volume $V$, and number of particles $N$, as is the case of the present system, the appropriate description is provided by the microcanonical ensemble. In this case, all accessible microstates compatible with the constraint $E_i=E$ are assumed to be equally probable. The probability distribution is uniform over the energy shell,
\begin{equation}
p_i =
\begin{cases}
\zeta(E)^{-1}, & E_i = E, \\
0, & E_i \neq E,
\end{cases}
\end{equation}
where $\zeta(E)$ denotes the number of accessible microstates at energy $E$. In the microcanonical ensemble, no partition function is introduced; instead, the central quantity is $\zeta(E)$, which enters thermodynamics through the definition of the entropy,
\begin{equation}
S(E,V,N) = k_B \ln \zeta(E).
\end{equation}

When the system is allowed to exchange energy with a thermal reservoir at fixed temperature $T$, while keeping $V$ and $N$ constant, the canonical ensemble applies \cite{r14,r12}. In this case, the probability of a microstate with energy $E_i$ is given by the Boltzmann distribution,
\begin{equation}
p_i = \frac{e^{-\tilde{\beta} E_i}}{Z},
\end{equation}
where $\tilde{\beta}=(k_B T)^{-1}$ and
\begin{equation}
Z(T,V,N) = \sum_i e^{-\tilde{\beta} E_i}
\end{equation}
is the canonical partition function \cite{r12}. The associated thermodynamic potential is the Helmholtz free energy,
\begin{equation}
F(T,V,N) = -k_B T \ln Z,
\end{equation}
which is the natural potential for systems at fixed temperature and volume. This is the ensemble suggested when a time perturbation to the boundary is introduced \cite{los1,los2,lenz,edlbuni}. In this case, depending on the set of control parameters and initial conditions, unbounded energy growth for the energy of the particles may be observed \cite{los1,los2}.

In our case, the energy is constant. Since there are no holes in the boundary, the particles can not escape through the border keeping the ensemble constant. We therefore seek for the probability distribution via the solution of the diffusion equation \cite{r18}.

The diffusion along the chaotic stripe can also be described via the solution of the diffusion equation. The probability of observing a particle with angle $\omega=\gamma^2-\gamma_{min}^2$ at a specific time $n$ is governed by ${{\partial P}\over{\partial n}}=D{{\partial^2P}\over{\partial\omega^2}}$ where $D$ is the diffusion coefficient obtained from \cite{liv1,liv2}
\begin{equation}
D=\lim_{n\to\infty}{{D_n}\over{n}},
\end{equation}
where
\begin{equation}
D_n=\sum_{i=1}^M<(\omega^i_n-\omega^i_0)^2>,
\end{equation}
where $M$ is the size of the ensemble of initial conditions. Since particles cannot cross the edges of the chaotic layer - at the price of viollating the Liouville's theorem \cite{r18}, the boundary conditions are 
\begin{equation}
\left.{{\partial P}\over{\partial\omega}}\right|_{\omega=(0,\omega_{fisc})}=0
\end{equation}
with the initial condition $P(\omega,0)=\delta(\omega-\omega_0)$. Here $\omega_{fisc}$ corresponds to the position of the upper curve bounding the chaotic diffusion. We use the technique of separation of variables \cite{butkov} to obtain the solution of the diffusion equation. Using both the boundary and the initial condition we obtain
\begin{equation}
P(\omega,n)={{1}\over{\omega_{fisc}}}+{{2}\over{\omega_{fisc}}}
e^{-{Dm^2\pi^2n}\over{\omega_{fisc}^2}}\left[b_m\cos\left({{m\pi\omega}\over{\omega_{fisc}}}\right)\right],
\end{equation}
with $m=1,2,3,\ldots$ where $b_m=\cos\left({{m\pi\omega_0}\over{\omega_{fisc}}}\right)$. The time-dependent term vanishes as $n\to\infty$ and corresponds to the transient dynamics. The stationary state is given by $P(\omega,n\to\infty)={{1}\over{\omega_{fisc}}}$. Therefore the existence of a curve $\omega_{fisc}$ as an upper border for the diffusion allows the determination of any moment of the probability distribution. The saturation of $\omega_{sat}$ is a consequence of the existence of $\omega_{fisc}$. This diffusion is only allowed due to the existence of a step length allowing the chaotic dynamics to evolve in the phase space. As discussed in Sec. \ref{sec4}, the order parameter goes to zero at the transition. It gives a separation of the regular dynamics observed for the integrable case when $\epsilon=0$ for the mixed dynamics when the control parameter $\epsilon\ne0$. Among of defining the transition, the control parameter $\epsilon$ also play the fundamental role of elementary excitation for the dynamics allowing the particles to diffuse along of the phase space.

\section{Topological defects}
\label{sec6}

This section is devoted to the discussion of topological defects present in phase space and their impact on particle transport. Inspecting Fig.~\ref{stripe}, we observe regions that remain empty, i.e., they are not visited by chaotic orbits. This behavior is a direct consequence of Liouville's theorem \cite{edlgeneris}: since attractors are absent in Hamiltonian systems, phase-space volume is preserved, and these empty regions correspond to stability islands associated with periodic orbits.

When chaotic trajectories approach such islands, they may become temporarily trapped in their vicinity \cite{r15,r16} before escaping and later being trapped again. These intermittent trapping events, combined with the presence of stability islands, break ergodicity and significantly modify the probability distribution of particles evolving within the chaotic layer. Nevertheless, because the chaotic stripe has a finite extent, bounded by the minimum and maximum values of chaotic motion [see Fig.~\ref{stripe}], the deviation around the average angle eventually saturates at long times.

As discussed in Ref.~\cite{r22}, the presence of periodic islands is dynamically equivalent to the existence of topological defects in phase space, which act as obstacles to transport and strongly influence the diffusive properties of the system.

There are several complementary ways to identify the presence of stickiness in dynamical systems \cite{return1,transport}. Stickiness is associated with the temporary trapping of trajectories wandering in the chaotic sea when they pass sufficiently close to regular structures in phase space, such as periodic islands or invariant spanning curves \cite{r15,r16}. In a predominantly chaotic regime, where such regular structures are absent or dynamically irrelevant, the return time of a trajectory to a given region of phase space typically follows an exponential distribution \cite{r27}, reflecting the absence of long-time correlations.

However, when periodic or quasiperiodic islands are present, they strongly influence the surrounding chaotic dynamics. Trajectories may remain trapped in their vicinity for long times before eventually returning to the chaotic sea. As a consequence, the exponential decay \cite{liv3} of the return-time distribution \cite{joelson}, characteristic of fully chaotic motion, is replaced at long times by a markedly slower decay, often of power-law type. This crossover signals the emergence of long trapping events and strong temporal correlations, and constitutes a well-established dynamical signature of stickiness.

\begin{figure}[t]
\centerline{(a)\includegraphics[width=0.8\linewidth]{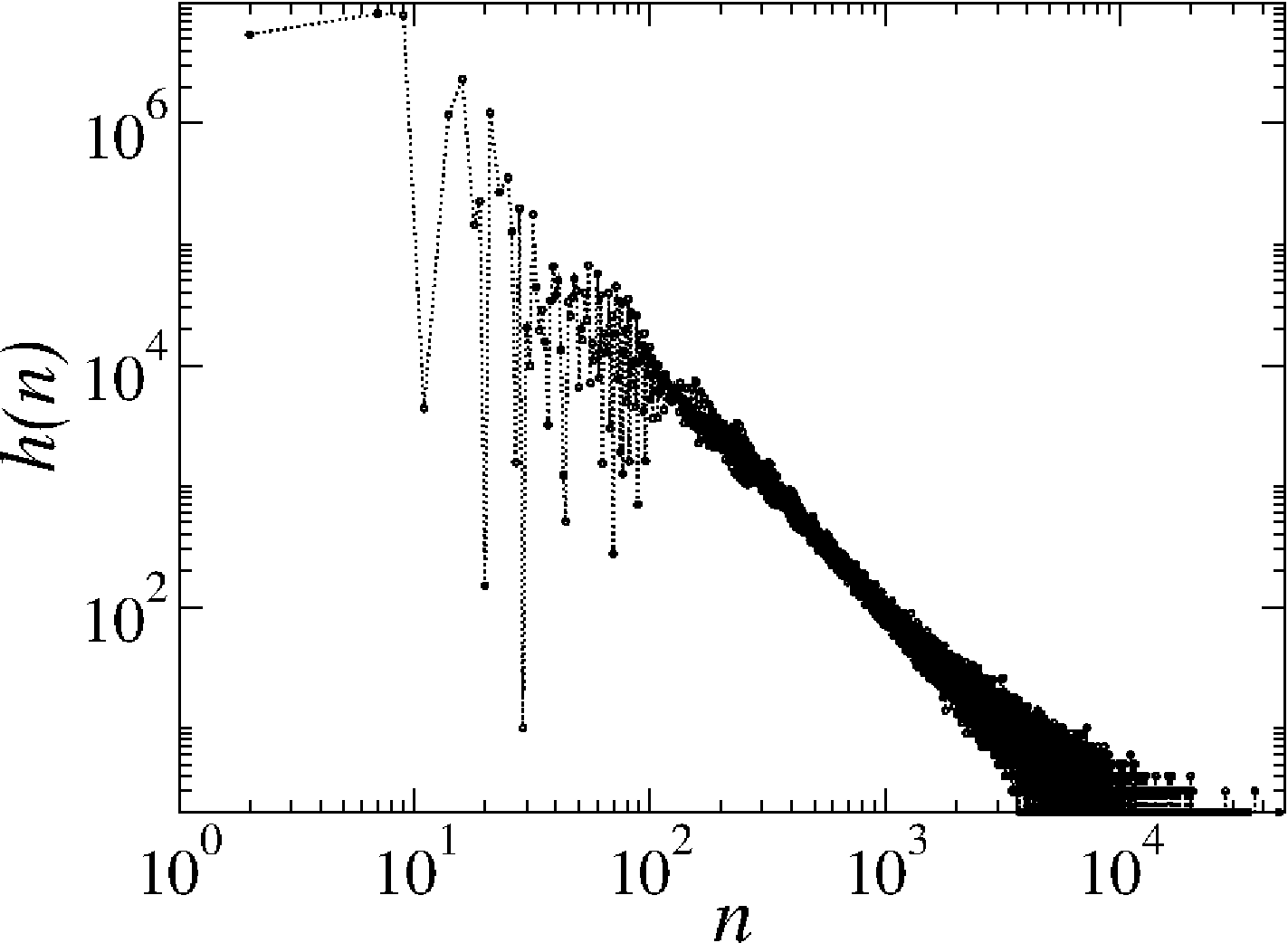}}
\centerline{(b)\includegraphics[width=0.8\linewidth]{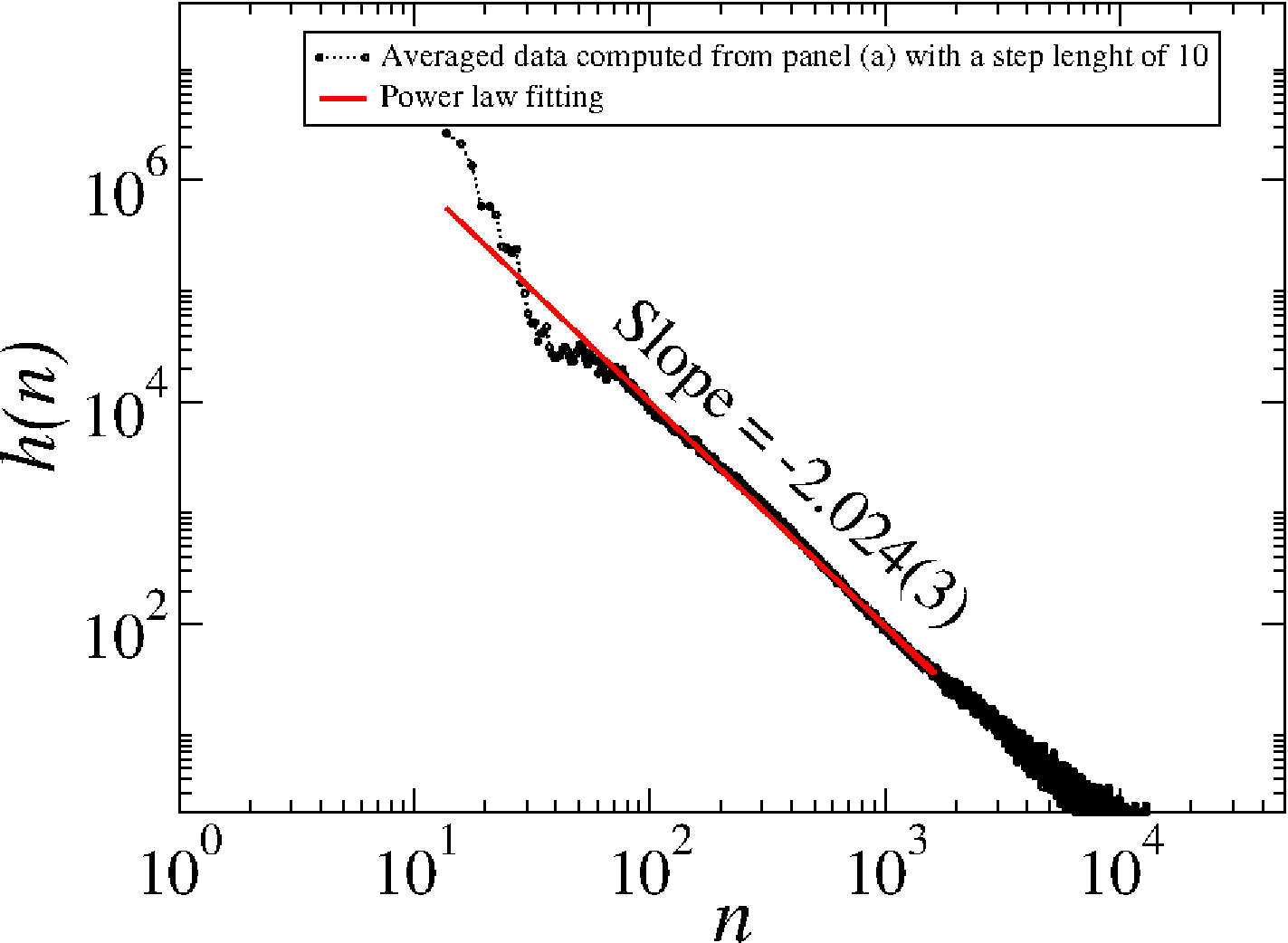}}
\caption{(a) Return-time distribution, $h(n)$ obtained from a single long orbit of $10^9$ collisions with the boundary, evolving inside a chaotic stripe and returning to a box of size $0.1$ centered at the saddle point $(\theta,\alpha)=(\pi,{{\pi}\over{2}})$. The control parameter is $\epsilon=0.05$. (b) Averaged data computed from panel (a) using a step length of $10$. A power-law fit yields a slope of $-2.024(3)$, confirming a slower decay when compared to the exponential behavior expected for a fully chaotic domain.}
\label{returning}
\end{figure}

Figure~\ref{returning}(a) shows the return-time distribution, $h(n)$ obtained from a single long orbit consisting of $10^9$ collisions with the boundary, for returns to a region defined by a square of side $0.1$ centered at the saddle point $(\theta,\alpha)=(\pi,{{\pi}\over{2}})$.

As shown in Fig.~\ref{returning}(a), some return times are forbidden due to the geometry of the boundary. For instance, a return at $n=1$ is not possible, since a particle initially located near $(\theta,\alpha)=(\pi,\pi/2)$ must first traverse the space and undergo a collision near $\theta\approx0$ before it can return to the original region. Similarly, due to the exponential separation of nearby trajectories, several small values of $n$ are not observed, such as $n=3,4,5,$ and $6$.

The statistics associated with short return times exhibit relatively large fluctuations when compared to those at longer times, for instance for $n>100$. By performing an averaging procedure with a step length of $10$, as shown in Fig.~\ref{returning}(b), the return-time distribution becomes smoother. A power-law fit to the decay region up to $n\approx1000$ collisions yields a slope of $-2.024(3)$, providing clear evidence of a power-law decay. This behavior confirms the presence of stickiness in the chaotic stripe, where chaotic motion develops along the phase space under the influence of nearby regular structures.

From a dynamical viewpoint, regular islands act as organizing centers that generate topological obstructions to transport in phase space. The stickiness induced by these structures can therefore be interpreted as a manifestation of topological defects embedded in an otherwise chaotic sea, whose presence is revealed through the anomalous statistics of return times.

\section{Discussions}
\label{sec7}

We have investigated the dynamics of an oval-like billiard \cite{r23,r5} with the aim of understanding a specific transition: from integrability ($\epsilon=0$) to non-integrability ($\epsilon\neq 0$). For $\epsilon=0$, the phase space is foliated \cite{r1} by parallel invariant lines, implying conservation of angular momentum which, together with energy conservation, provides the two constants of motion required for integrability in a system with two degrees of freedom \cite{r24}. When $\epsilon\neq 0$, the phase-space structure changes qualitatively and a chaotic layer emerges. As the control parameter increases, this chaotic layer grows until all invariant spanning curves responsible for whispering-gallery orbits \cite{r5} are destroyed.

Diffusion along the resulting chaotic stripe is finite and scales with the control parameter according to $\omega_{rms,{\rm sat}}\propto\epsilon^{\tilde{\alpha}}$, with $0<\tilde{\alpha}<1$, where we measure $\tilde{\alpha}=0.507(2)$. In this framework, the order parameter $\omega_{rms,{\rm sat}}$ vanishes continuously at the transition, while its susceptibility diverges, in agreement with the phenomenology expected for second-order (continuous) phase transitions \cite{r10,r22,r15}.

The elementary excitation \cite{r22} responsible for driving the diffusive dynamics is the control parameter $\epsilon$ itself, which deforms the boundary and governs the onset and growth of chaotic diffusion in phase space. In addition, topological defects associated with stability islands affect transport through local trapping events, commonly referred to as stickiness \cite{r22,r16,r15}, thereby modifying the statistical properties of diffusion.

\begin{figure}[htb]
\centerline{\includegraphics[width=1.0\linewidth]{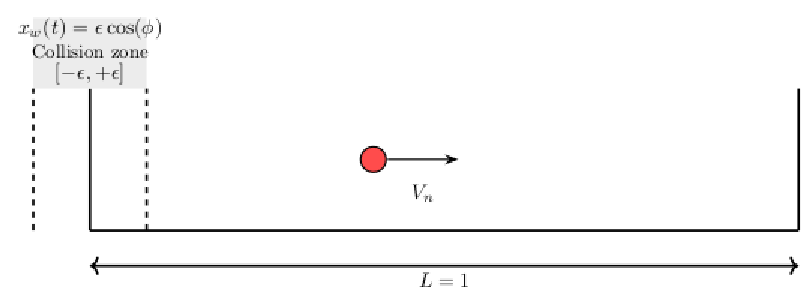}}
\caption{Sketch of the Fermi--Ulam model.}
\label{fum}
\end{figure}

Let us now discuss how the present results fall into universality classes previously reported in the literature. We start with the Fermi--Ulam model \cite{lich}. The system consists of a classical particle confined to move between two walls. One wall is fixed at position $l$, while the other oscillates around an equilibrium position according to
$x_w(t)=\epsilon\cos(\omega t)$, where $\omega$ is the angular frequency and $\epsilon$ is the oscillation amplitude. Figure~\ref{fum} shows a sketch of the model.

Using dimensionless variables, the dynamics of the system is described by a two--dimensional map relating the particle velocity and the phase of the moving wall. The mapping connects the $n$th collision to the subsequent $(n+1)$th collision and is given by
\begin{equation}
\left\{\begin{array}{ll}
\phi_{n+1}=[\phi_n+{{2}\over{V_n}}]~{\rm mod}~(2\pi),\\
V_{n+1}=|V_n-2\epsilon\sin(\phi_{n+1})|.\\
\end{array}
\right.
\label{c10_eq22}
\end{equation}

The mapping~(\ref{c10_eq22}) contains a control parameter $\epsilon$ that governs the transition from integrability to non--integrability. When $\epsilon=0$, the system is integrable and the particle velocity remains constant. For $\epsilon\neq0$, the phase space acquires a mixed structure, with the emergence of a chaotic sea in the low--energy region. This chaotic domain is bounded by a set of invariant spanning curves, which act as barriers in phase space and prevent trajectories from crossing them.

Such barriers do not violate Liouville's theorem, since the determinant of the Jacobian matrix associated with mapping~(\ref{c10_eq22}) is unity, ensuring area preservation. The position \cite{r8} of the lowest invariant spanning curve scales with the control parameter as $\epsilon^{\tilde{\alpha}}$, with critical exponent $\tilde{\alpha}=0.5$. Consequently, the size of the chaotic domain in the low-energy regime is limited by $\epsilon^{0.5}$.

When an ensemble of initial conditions is launched inside the chaotic sea at low energies, the particles undergo diffusive dynamics. For a sufficiently large ensemble, the dispersion initially grows according to a power law, which can be written as $\tilde{\omega}\propto (n\epsilon)^{\beta}$, with $\beta=1/2$ and $\tilde{\omega}$ denotes the dispersion of chaotic orbits. This growth regime, however, is not persistent. After a characteristic number of collisions with the moving wall, denoted by $n_x$, the power-law increase is replaced by a saturation plateau, which scales with $\tilde{\omega}_{sat}\propto\epsilon^{\alpha}$ with $\alpha=1/2$.

The crossover from the growth regime to saturation is governed by the scaling relation $n_x\propto\epsilon^{z}$, where both $\beta$ and $z$ play the role of critical exponents. In the Fermi-Ulam model~\cite{r8}, these exponents were found to be $\beta=1/2$ and $z=-1$, characterizing the transition between the diffusive and saturated regimes.

Physically, the power-law growth of the dispersion indicates that the system has not yet reached its correlation length, and long-range correlations still dominate the dynamics. In contrast, the emergence of a constant plateau signals that the correlation length has been fully explored. At this stage, correlations are effectively lost, and the system reaches a stationary regime. The saturation of the dispersion therefore provides a clear dynamical signature of the underlying transition, analogous to the behavior observed near critical points in phase transitions.

The scenario observed in the Fermi-Ulam model \cite{r8} is exactly the same as that found in the model discussed in the present work: a lower bound for chaotic motion imposed by the minimal position of the moving wall (e.g., zero), and an upper bound provided by the first invariant spanning curve. The chaotic dispersion in the Fermi-Ulam model exhibits a saturation value \cite{r8} that scales as $\epsilon^{0.5}$, indicating that a stationary probability distribution is reached. Remarkably, the critical exponent observed here, $\tilde{\alpha}=0.507(2)$, coincides with that reported for the Fermi--Ulam model \cite{r8}.

We now turn to another system in which time does not explicitly appear: a periodically corrugated waveguide \cite{r9,bliock}. The model consists of a classical light ray undergoing specular reflections between a corrugated surface described by $y=y_0+d\cos(kx)$ and a flat surface located at $y=0$. Here, $y_0$ denotes the average distance between the two surfaces, $d$ is the corrugation amplitude, and $k$ is the wave number. The dynamical variables are the angle $\theta$ of the ray trajectory measured from the positive horizontal axis, and the corresponding $x$ coordinate at the instant of reflection.

\begin{figure}[htb]
\centerline{(a)\includegraphics[width=0.475\linewidth]{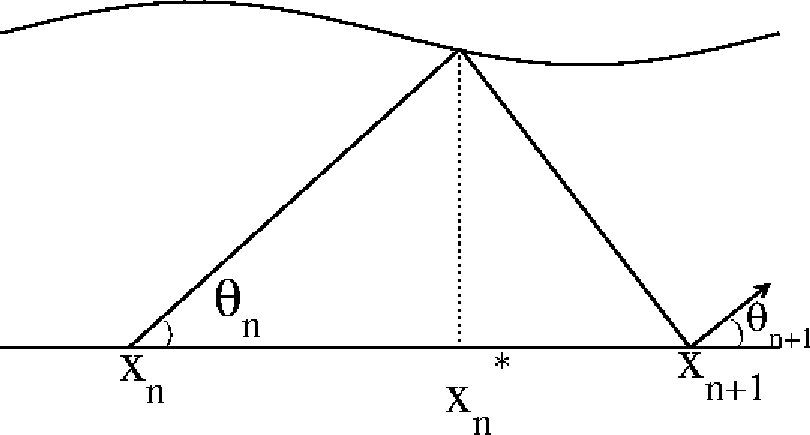}
(b)\includegraphics[width=0.475\linewidth]{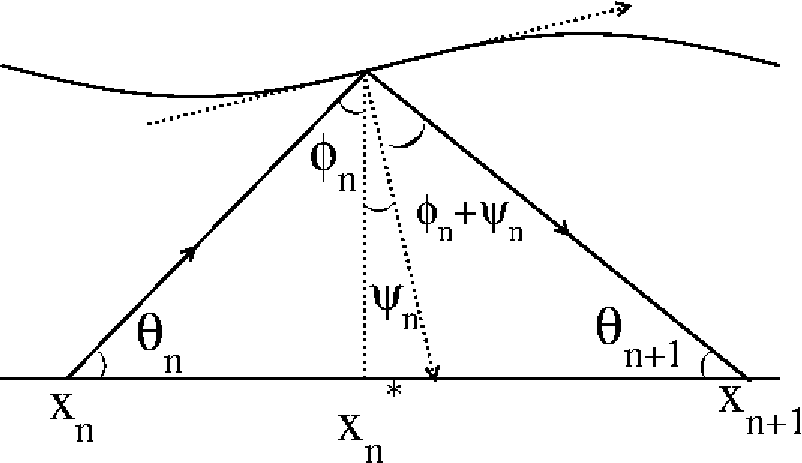}
}
\caption{(a) Reflection of a light ray from the corrugated surface after leaving the flat surface at $y=0$. (b) Details of the trajectory before and after a collision with the corrugated surface.}
\label{waveguide}
\end{figure}

The mapping is iterated whenever the ray hits the flat surface at $y=0$; therefore, multiple reflections on the corrugated surface are neglected. A discussion when multiple collisions can be seen in Ref. \cite{mariojpa} Figure~\ref{waveguide} shows a sketch of the model.

Using dimensionless variables \cite{r9}, the dynamics of the light ray is governed by the following mapping:
\begin{equation}
\left\{\begin{array}{ll}
X_{n+1}=X_n+\left[{{1}\over{\gamma_n}}+{{1}\over{\gamma_{n+1}}}\right]~
{\rm mod}~(2\pi),\\
\gamma_{n+1}=\gamma_n+2\delta\sin\left(X_n+{{1}\over{\gamma_n}}\right).
\end{array}
\right.
\label{wave_map}
\end{equation}

The mapping~(\ref{wave_map}) is also area preserving, since the determinant of its Jacobian matrix is equal to unity. The dispersion of chaotic trajectories in phase space follows a scenario similar to that of the Fermi-Ulam model \cite{r8}, with the presence of an invariant spanning curve acting as a barrier that prevents trajectories from crossing it.

The control parameter $\delta$ governs the transition from integrability to non-integrability. When $\delta=0$, the phase space is foliated by invariant curves, since the angle $\gamma_n$ remains constant. For any $\delta\neq0$, a mixed phase-space structure emerges, allowing a chaotic sea to develop until it reaches a limiting invariant barrier, whose position \cite{r9} scales as $\delta^{0.5}$.

Although the periodically corrugated waveguide differs from the Fermi--Ulam model in its physical realization, both systems exhibit identical behavior near the transition from integrability to non--integrability, sharing the same critical scaling properties. We therefore conclude that, close to the transition, both systems belong to the same universality class.

A closely related scenario has been reported for a family of area-preserving mappings \cite{r10} of the form
\begin{equation}
\left\{\begin{array}{ll}
I_{n+1}=I_n+\epsilon \sin(\theta_n),\\
\theta_{n+1}=\left[\theta_n+\dfrac{1}{|I_{n+1}|^{\tilde{\gamma}}}\right]\ {\rm mod}\ (2\pi),
\end{array}
\right.
\label{eq1}
\end{equation}
with $\tilde{\gamma}>0$. In these systems, diffusive behavior saturates due to the presence of invariant spanning curves in phase space, whose location can be estimated as
\begin{equation}
I_{fisc}\cong \left[\dfrac{\tilde{\gamma}\epsilon}{0.9716\ldots}\right]^{1/(1+\tilde{\gamma})}.
\end{equation}

The case $\tilde{\gamma}=1$ recovers both the Fermi-Ulam model \cite{r8} and the periodically corrugated waveguide \cite{r9,bliock}. The critical exponent governing the diffusive behavior in these systems \cite{r10,r22} is given by $\alpha=1/(1+\tilde{\gamma})$. Consequently, for $\tilde{\gamma}=1$, one finds $\alpha=1/2$, remarkably the same value observed for the transition reported in the present work.

Therefore, despite the distinct physical realizations, ranging from the Fermi-Ulam model \cite{r8} and periodically corrugated waveguides \cite{r9} to the family of area-preserving maps \cite{r10} defined by Eq.~(\ref{eq1}) for $\tilde{\gamma}=1$, the transition from integrability to non-integrability observed here belongs to the same universality class as that of the Fermi-Ulam model.

\section{Investigations for other models: a setup}
\label{sec8}

As discussed throughout this paper, the transition from integrability to non-integrability can be systematically characterised by a well-defined sequence of steps. These steps are guided by four fundamental questions, which we propose here as a general framework for future investigations of dynamical phase transitions in conservative systems:

(1) \emph{Symmetry breaking}: What is the symmetry that is broken at the transition? Is the symmetry breaking manifested in phase space, or does it arise at a more fundamental level, such as in the algebraic structure of the equations of motion?

(2) \emph{Order parameter}: What observable plays the role of an order parameter for the transition? Does it vanish continuously at criticality? How does its associated susceptibility behave near the transition? Is there a divergence, and if so, what are the corresponding scaling laws?

(3) \emph{Elementary excitation}: In the context of diffusive dynamics, what is the elementary excitation that enables transport in phase space? Which control parameter or dynamical mechanism governs the onset of diffusion?

(4) \emph{Topological defects}: Are there topological structures in phase space that affect transport properties? Do they induce stickiness through temporary trapping near stability islands, or do they act via other mechanisms thereby breaking ergodicity?

Within this framework, we propose a set of open problems that may be addressed by following the sequence of steps outlined above. We begin with perturbations of the circle. If the circle is cut along a diameter and the two semicircles are separated and connected by straight segments, one obtains the classical Bunimovich stadium \cite{r2,r4}. Another geometrical variation is the lemon billiard \cite{lemon}, constructed as the intersection of two identical circles.

\bigskip
\noindent
{\bf Problem 1 (Lemon billiard).}  
The lemon billiard \cite{lemon} is defined by a boundary formed from the intersection of two identical circles of radius $R$, whose centres are separated by a distance $2B$, with $2B<2R$. The circles are positioned symmetrically along the horizontal axis, with centres located at $x=\pm B$, where $B\in[0,R)$. In this system, it is natural to expect the existence of two distinct transitions: one at $B=0$, corresponding to the circular billiard, and another as $B$ approaches $R$, where the geometry becomes singular. How does the phase transition behave in these regimes? Are these transitions continuous or discontinuous, and do they share common scaling properties or universality classes?

\begin{figure}[htb]
\centerline{\includegraphics[width=0.6\linewidth]{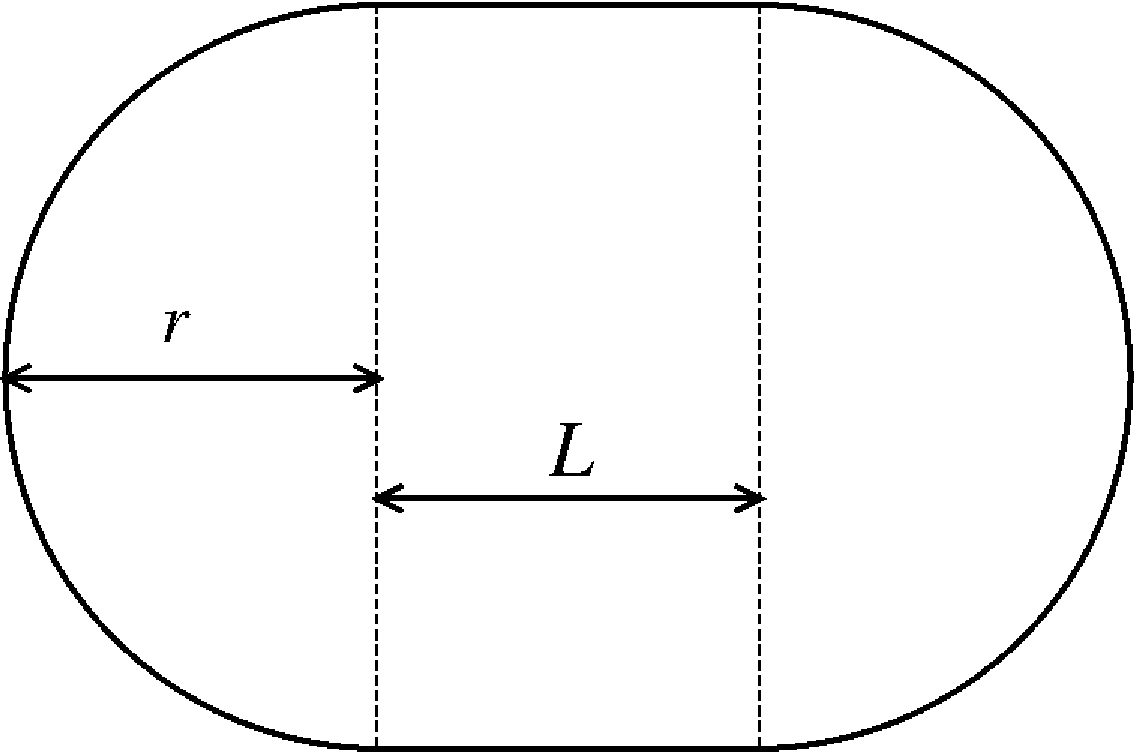}}
\caption{Scketch of the stadium billiard which is formed by two semicircles of radius $r$ connected by two parallel straight segments of length $L$.}
\label{stadium}
\end{figure}

\bigskip
\noindent
{\bf Problem 2 (Bunimovich stadium billiard).}  
The Bunimovich stadium billiard \cite{r2,r4} consists of a planar domain formed by two semicircles of radius $r$ connected by two parallel straight segments of length $L$. A point-like particle moves freely inside the domain with constant speed and undergoes specular (elastic) reflections at the boundary. Figure \ref{stadium} shows a scketch of the Bunimovich billiard.

Between collisions, the particle dynamics is governed by free motion,
\begin{equation}
\dot{\mathbf{r}} = \mathbf{v}, \qquad \dot{\mathbf{v}} = \mathbf{0},
\end{equation}
where $\mathbf{r}=(x,y)$ is the position vector and $\mathbf{v}$ is the velocity. At collisions with the boundary, the velocity is updated according to the specular reflection law,
\begin{equation}
\mathbf{v}^+ = \mathbf{v}^- - 2(\mathbf{v}^- \cdot \mathbf{n})\,\mathbf{n},
\end{equation}
where $\mathbf{n}$ is the outward unit normal vector at the collision point, and the superscripts $-$ and $+$ denote before and  after collision velocities, respectively.

Despite the presence of flat boundary segments, the stadium billiard exhibits global hyperbolic dynamics \cite{r2}. The chaotic behavior originates from the defocusing mechanism introduced by Bunimovich \cite{r2}, whereby trajectories focused by the curved boundary components are subsequently defocused after reflections on the straight segments. As a consequence, the stadium billiard is ergodic, mixing, and possesses positive Lyapunov exponents for almost all initial conditions.

The central question is the following: at the transition from integrability (the circular billiard limit) to non-integrability, does this transition belong to a known universality class? Does it share common order parameters or susceptibilities with the oval-like billiard studied in the present work?

\bigskip
\noindent
{\bf Problem 3 (Balescu tokamap).}  
The Balescu tokamap \cite{balescu} provides a simplified, area-preserving description of magnetic field-line dynamics in the presence of perturbations and captures the transition from regular to chaotic motion as the perturbation strength increases. In its original formulation, the map is written as
\begin{align}
K_{n+1} &= K_n - \frac{L}{2\pi}\,
\frac{\sin(2\pi T_n)}{1 + K_n}, \label{eq:tokamap_K} \\[4pt]
T_{n+1} &= T_n + \frac{1}{q(K_{n+1})}
- \frac{L}{2\pi}\,
\frac{\cos(2\pi T_n)}{(1 + K_n)^2},
\label{eq:tokamap_T}
\end{align}
where $K_n$ is a radial-like action variable associated with the magnetic surface label, $T_n$ is an angular variable defined modulo unity, $L$ is the control parameter measuring the strength of the magnetic perturbation, and $q(K)$ denotes the safety-factor profile of the tokamak.

For vanishing perturbation ($L=0$), the map is integrable and trajectories lie on invariant tori characterised by the rotation number $1/q(K)$. As $L$ increases, nonlinear resonances emerge, giving rise to island chains and stochastic layers. For sufficiently large values of $L$, invariant tori are progressively destroyed and extended chaotic regions develop in phase space, enabling enhanced radial transport of magnetic field lines. The tokamap is symplectic (area preserving) and therefore constitutes a discrete Hamiltonian system. As such, it provides a minimal yet powerful framework for investigating the onset of chaos, resonance overlap, and transport processes in magnetically confined plasmas.

The central question is whether the phase transition from integrability to non-integrability, governed by the control parameter $L$, can be characterised within the framework developed in the present paper. In particular, does this transition admit an order parameter, a diverging susceptibility, and well-defined critical exponents? If so, does it belong to a known universality class?

\bigskip
\noindent
{\bf Problem 4 (Standard nontwist map).}  
The standard nontwist map \cite{sntm} is a paradigmatic area-preserving discrete-time system arising in plasma physics \cite{plasma}, particularly in configurations where the twist condition is violated due to a non-monotonic rotation profile. Unlike the standard map \cite{lich}, the rotation number in the nontwist map exhibits an extremum, leading to qualitatively distinct dynamical behavior.

A commonly studied form of the standard nontwist map is given by
\begin{align}
x_{n+1} &= x_n - b \sin(2\pi y_n), \label{eq:nontwist_x} \\[4pt]
y_{n+1} &= y_n + a\bigl(1 - x_{n+1}^2\bigr), \label{eq:nontwist_y}
\end{align}
where $x_n$ and $y_n$ are phase-space variables defined modulo unity, and $a$ and $b$ are control parameters. The violation of the twist condition occurs at $x=0$, where the shear vanishes.

For small values of the perturbation parameter $b$, the dynamics is predominantly regular, with trajectories confined to invariant tori. As $b$ increases, resonant island chains, meandering invariant curves, and chaotic layers emerge, leading to a mixed phase-space structure. For sufficiently large perturbations, global chaos sets in due to the destruction of invariant tori, resulting in enhanced transport across the shearless region. Owing to its symplectic nature, the standard nontwist map provides a minimal model for studying shearless transport barriers, reconnection scenarios, and the transition from regular to chaotic dynamics in Hamiltonian systems without twist.

Within the framework proposed in this paper, a natural question is how the phase transition from integrability to non-integrability develops as the control parameter $b$ is varied from $b=0$ to $b\neq 0$. Are there critical exponents governing this transition, and can they be used to define universality classes analogous to those identified in the oval-like billiard?

\bigskip
\noindent
{\bf Problem 5 (Circular restricted three-body problem).}  
We briefly describe the planar circular restricted three-body problem (CR3BP) \cite{prtc1,prtc2} in the uniformly rotating (synodic) frame and define the Poincar\'e section used to construct the associated map. In this problem, two primaries with masses $m_1$ and $m_2$ move on circular orbits about their common centre of mass, while a third particle of negligible mass evolves under their gravitational attraction.

After the standard nondimensionalisation (total mass $m_1+m_2=1$, distance between the primaries equal to unity, and angular velocity of the rotating frame equal to one), the dynamics depends on a single parameter,
\begin{equation}
\mu = \frac{m_2}{m_1+m_2},
\end{equation}
which measures the mass ratio of the primaries. In this formulation, $\mu$ controls the departure from the integrable Keplerian limit: as $\mu$ increases from zero, resonances and chaotic layers progressively develop in phase space.

Let $(x,y)$ denote the coordinates of the massless particle in the rotating frame. The primaries are fixed on the $x$ axis at
\begin{equation}
(x_1,y_1)=(-\mu,0), \qquad (x_2,y_2)=(1-\mu,0).
\end{equation}
The distances to the primaries are defined as
\begin{equation}
r_1=\sqrt{(x+\mu)^2+y^2}, \qquad 
r_2=\sqrt{(x-1+\mu)^2+y^2}.
\end{equation}
The effective potential takes the form
\begin{equation}
V(x,y)=\frac{1}{2}(x^2+y^2)+\frac{1-\mu}{r_1}+\frac{\mu}{r_2},
\end{equation}
and the equations of motion are
\begin{align}
\ddot{x}-2\dot{y} &= \frac{\partial V}{\partial x}
= x-\frac{(1-\mu)(x+\mu)}{r_1^{3}}
-\frac{\mu(x-1+\mu)}{r_2^{3}}, \label{eq:cr3bp_x}\\
\ddot{y}+2\dot{x} &= \frac{\partial V}{\partial y}
= y-\frac{(1-\mu)y}{r_1^{3}}
-\frac{\mu y}{r_2^{3}}. \label{eq:cr3bp_y}
\end{align}

An additional integral of motion is provided by the Jacobi constant,
\begin{equation}
C = 2V(x,y)-\left(\dot{x}^2+\dot{y}^2\right),
\label{eq:jacobi}
\end{equation}
which constrains the admissible region of motion through the condition $2V(x,y)\ge \dot{x}^2+\dot{y}^2$.

To visualise the organisation of phase space and construct a discrete-time description, we introduce a standard Poincar\'e section exploiting the symmetry with respect to the line joining the primaries. Specifically, we define
\begin{equation}
\Sigma = \left\{(x,y,\dot{x},\dot{y})\in\mathbb{R}^4 
\ \middle|\ y=0,\ \dot{y}>0\right\}.
\label{eq:poincare_section}
\end{equation}
Restricting the flow to crossings with $\dot{y}>0$ avoids double counting and yields a well-defined return map on $\Sigma$.

For fixed values of $(\mu,C)$, the Jacobi integral~\eqref{eq:jacobi} determines $\dot{y}$ at each intersection,
\begin{equation}
\dot{y} = \left[\,2V(x,0)-C-\dot{x}^2\,\right]^{1/2},
\qquad \dot{y}>0,
\label{eq:ydot_on_section}
\end{equation}
so that the Poincar\'e map can be represented as an area-preserving two-dimensional map,
\begin{equation}
(x_n,\dot{x}_n)\ \mapsto\ (x_{n+1},\dot{x}_{n+1}),
\label{eq:poincare_map}
\end{equation}
generated by successive intersections of a trajectory with $\Sigma$.

The central question is how chaos develops in the CR3BP as the control parameter $\mu$ is varied. In particular, can the transition from the integrable limit $\mu\to0$ to mixed and chaotic dynamics be characterised within the framework proposed in this paper? Does this transition exhibit scaling laws, critical exponents, or belong to a well-defined universality class?

\section{Conclusions}
\label{sec9}

In this work we have investigated, in a systematic and quantitative way, a transition from integrability to non-integrability in an oval-like billiard with static boundary \cite{r23,r5}. By varying a single control parameter $\epsilon$, which deforms the boundary from a circular shape to a noncircular one, the system undergoes a qualitative change in its phase-space structure, evolving from a completely regular dynamics \cite{r1} to a mixed phase space \cite{r23} where chaotic diffusion becomes possible.

We have shown that this transition exhibits all the hallmark features of a continuous \cite{r11,r12} (second-order) dynamical phase transition \cite{r14,r13}. In particular, the destruction of angular-momentum conservation constitutes the symmetry breaking that separates the integrable from non-integrable regimes. The emergence of a chaotic stripe in phase space allows for diffusive motion, whose extent is quantified by an order parameter \cite{r22}, $\omega_{rms,{\rm sat}}$, defined from the saturation of the diffusive spreading of trajectories. This order parameter vanishes continuously as $\epsilon\to0$, while its susceptibility diverges, revealing critical behavior governed by a well-defined scaling law.

The control parameter $\epsilon$ plays a dual role in this scenario. Besides driving the transition, it also acts as the elementary excitation \cite{r17} responsible for enabling diffusion in phase space. This interpretation provides a clear dynamical meaning to the onset of chaos and connects the present results to broader concepts of transport and excitation in nonlinear systems. Furthermore, the presence of stability islands embedded in the chaotic sea \cite{r16,r15} introduces topological defects in phase space, leading to stickiness \cite{r17} and local trapping events that modify transport properties \cite{joelson} without suppressing the overall diffusive behavior.

A central result of this study is the identification of a critical exponent $\tilde{\alpha}\approx1/2$, which governs the scaling of the order parameter near the transition. Remarkably, the same exponent appears in a wide class of seemingly unrelated conservative systems, including the Fermi-Ulam model \cite{r8,lich}, periodically corrugated waveguides \cite{r9,bliock}, and families of area-preserving mappings \cite{r10} (for the case of $\gamma=1$). This strongly supports the conclusion that the transition reported here belongs to a broader universality class of integrability-breaking transitions in Hamiltonian dynamics.

Beyond the specific billiard considered, we have proposed a general framework for investigating dynamical phase transitions in conservative systems. This framework is based on four key ingredients: the identification of symmetry breaking, the definition of an appropriate order parameter and its susceptibility, the determination of the elementary excitation driving diffusion, and the role of topological defects in phase space. We have illustrated how this approach can be naturally extended to a variety of paradigmatic models, including the lemon \cite{lemon} billiard, the Bunimovich stadium \cite{r2,r4}, the Balescu tokamap \cite{balescu}, the standard nontwist map \cite{sntm}, and the circular restricted three-body problem \cite{prtc1,prtc2}.

We believe that the perspective developed here provides a unifying view of integrability-breaking phenomena in nonlinear dynamics, bridging concepts from chaos theory \cite{hilborn}, statistical mechanics \cite{r11,r12,r13,r17,r18}, and Hamiltonian transport \cite{r19,r20}. The results presented in this work not only deepen the understanding of billiard dynamics \cite{r1} but also open new avenues for the systematic classification of dynamical phase transitions \cite{r13,r22} and their universality classes across a wide range of physical systems.

\section*{Ackowledgments}

E.D.L. acknowledges support from Brazilian agencies CNPq (No. 301318/2019-0, 304398/2023-3) and FAPESP (No. 2019/14038-6 and No. 2021/09519-5)

\appendix
\section{Lyapunov exponent algorithm}
\label{Aapp1}

In this Appendix we discuss an algorithm to compute the Lyapunov exponent \cite{eckman}. We start by presenting the procedure for one-dimensional systems and then extend it to two-dimensional mappings. Consider the dynamics of a one-dimensional map given by $x_{n+1}=f(x_n)$, where $f$ is a nonlinear function. The distance between two nearby initial conditions after $n$ iterations is defined as
\begin{equation}
d=|f^{(n)}(x_0+\varepsilon)-f^{(n)}(x_0)|,
\label{c8_eq14}
\end{equation}
where $\varepsilon$ is arbitrarily small and $x_0$ denotes the initial condition. The relative distance is therefore given by $d/\varepsilon$. Assuming that this quantity grows exponentially with $n$, one writes ${{d}\over{\varepsilon}}=e^{\lambda n}$, where $\lambda$ is the Lyapunov exponent. Taking the limit $\varepsilon\rightarrow 0$, we obtain
\begin{equation}
\lim_{\varepsilon\rightarrow 
0}\Big\vert{{f^{(n)}(x_0+\varepsilon)-f^{(n)}(x_0)}\over{\varepsilon}}\Big\vert=e^{
\lambda n},
\label{c8_eq15}
\end{equation}
which leads to $|f^{\prime (n)}(x_0)|=e^{\lambda n}$. Isolating $\lambda$, we find $\lambda n=\ln|f^{\prime (n)}(x_0)|$. Using the chain rule, this expression can be written as
\begin{equation}
\ln|f^{\prime 
(n)}(x_0)|=\ln|f^{\prime}(x_{n-1})f^{\prime}(x_{n-2})f^{\prime}(x_{n-3})\ldots 
f^{\prime}(x_{0})|.
\label{eq_new}
\end{equation}

For the two-dimensional case where the dynamics is described by
\begin{eqnarray}
x_{n+1}&=&F(x_n,y_n),\nonumber\\
y_{n+1}&=&G(x_n,y_n),\nonumber
\end{eqnarray}
where $F$ and $G$ are nonlinear functions of their variables, the Lyapunov exponents are defined as
\begin{equation}
\lambda_j=\lim_{n\rightarrow\infty}{{1}\over{n}}\ln|\Lambda_n^{(j)}|,
\label{c10_eq20}
\end{equation}
with $j=1,2$, where $\Lambda_n^{(j)}$ denote the eigenvalues of the Jacobian matrix
$M=\Pi_{i=1}^nJ_i(x_i,y_i)=J_nJ_{n-1}J_{n-2}\ldots J_2J_1$.
The Jacobian matrix is written as
$$
J=\left(\begin{array}{ll}
{{\partial F}\over{\partial x_n}}  &  {{\partial F}\over{\partial y_n}}  \\
{{\partial G}\over{\partial x_n}}  &  {{\partial G}\over{\partial y_n}}\\
\end{array}
\right),
$$

Since convergence is achieved only for large values of $n$, the direct multiplication of the matrices $J_i$ may lead to numerical overflow, making the direct computation of $\lambda$ impractical.

To overcome this difficulty, we employ an algorithm proposed by Eckmann and Ruelle \cite{eckman}, which allows the Jacobian matrix to be written as a product $J=\Theta T$, where $\Theta$ is an orthogonal matrix (for which the transpose equals the inverse, i.e., $\Theta^{-1}=\Theta^t$) and $T$ is an upper triangular matrix. Explicitly,
$$
\Theta=\left(\begin{array}{ll}
\cos(\theta)  &  -\sin(\theta)  \\
\sin(\theta)  &    \cos(\theta)\\
\end{array}
\right),
$$
and
$$
T=\left(\begin{array}{ll}
T_{11}  &  T_{12}  \\
0  &    T_{22}\\
\end{array}
\right).
$$

Using this decomposition, the matrix $M$ can be written as
\begin{eqnarray}
M&=&J_nJ_{n-1}J_{n-2}\ldots J_2J_1,\nonumber\\
&=&J_nJ_{n-1}J_{n-2}\ldots J_2\Theta_1\Theta_1^{-1}J_1.
\end{eqnarray}
Defining $T_1=\Theta_1^{-1}J_1$ and $\tilde{J}_2=J_2\Theta_1$, the elements of the matrix $T_1$ are given by
$$
\left(\begin{array}{ll}
T_{11}  &  T_{12}  \\
0  &    T_{22}\\
\end{array}
\right)=
\left(\begin{array}{ll}
\cos(\theta)  &  \sin(\theta)  \\
-\sin(\theta)  &   \cos(\theta)\\
\end{array}
\right)
\left(\begin{array}{ll}
j_{11}  &  j_{12}  \\
j_{21}  &    j_{22}\\
\end{array}
\right).
$$
Imposing the condition $T_{21}=0$ yields
$0=-j_{11}\sin(\theta)+j_{21}\cos(\theta)$, which leads to
\begin{equation}
{{j_{21}}\over{j_{11}}}={{\sin(\theta)}\over{\cos(\theta)}}.
\label{c10_eq21}
\end{equation}
Instead of computing $\theta=\arctan(j_{21}/j_{11})$ numerically, which is computationally expensive, we obtain $\sin(\theta)$ and $\cos(\theta)$ directly from the elements of $J$, leading to
\begin{eqnarray}
\cos(\theta)&=&{{j_{11}}\over{\sqrt{j_{11}^2+j_{21}^2}}},\\
\sin(\theta)&=&{{j_{21}}\over{\sqrt{j_{11}^2+j_{21}^2}}}.
\end{eqnarray}

The expressions for $T_{11}$ and $T_{22}$ then follow as
$T_{11}=j_{11}\cos(\theta)+j_{21}\sin(\theta)$ and
$T_{22}=-j_{12}\sin(\theta)+j_{22}\cos(\theta)$, yielding
\begin{eqnarray}
T_{11}&=&{{j_{11}^2+j_{21}^2}\over{\sqrt{j_{11}^2+j_{21}^2}}},\\
T_{22}&=&{{j_{11}j_{22}-j_{12}j_{21}}\over{\sqrt{j_{11}^2+j_{21}^2}}}.
\end{eqnarray}

Once the elements $T_{11}$ and $T_{22}$ are obtained, the transformed matrix $\tilde{J}_2$ is computed as $\tilde{J}_2=J_2\Theta_1$, that is,
$$
\left(\begin{array}{ll}
\tilde{j}_{11}  &  \tilde{j}_{12}  \\
\tilde{j}_{21}  &  \tilde{j}_{22}\\
\end{array}
\right)=
\left(\begin{array}{ll}
j_{11}  &  j_{12}  \\
j_{21}  &  j_{22}\\
\end{array}
\right)
\left(\begin{array}{ll}
\cos(\theta)  &  -\sin(\theta)  \\
\sin(\theta)  &  \cos(\theta)\\
\end{array}
\right).
$$
This procedure is repeated iteratively until the full sequence of Jacobian matrices $J_n$, $J_{n-1}$, $J_{n-2}$, and so on, is exhausted. The Lyapunov exponents are then obtained as
\begin{equation}
\lambda_{j}=\lim_{n\rightarrow\infty}\sum_{i=1}^n\ln|T_{jj}^{(i)}|,~j=1,
2.
\end{equation}

\begin{figure}[t]
\centerline{\includegraphics[width=1.0\linewidth]{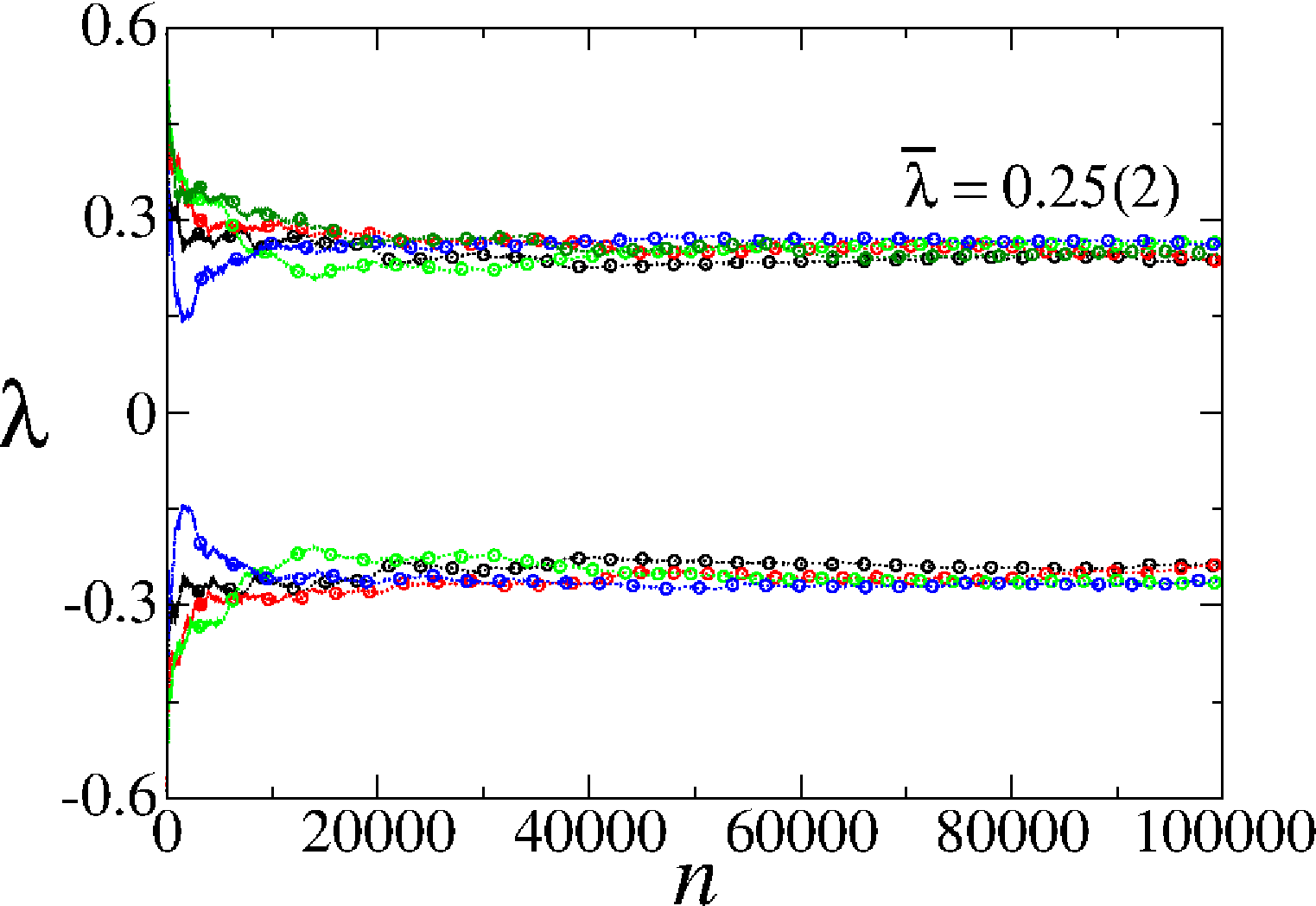}}
\caption{(Color online) Plot of the positive and negative Lyapunov exponent for the billiard model described by mapping (\ref{B_eq4}). We considered the control parameter $\epsilon=0.05$ and chose five different initial conditions distribuited randomly along the chaotic sea. The convergence of the positive Lyapunov exponent for the five sample is $\bar{\lambda}=0.25(2)$. Because the system is Hamiltonian, the positive and negative expoents must be symmetric.}
\label{app1}
\end{figure}

Figure \ref{app1} shows the convergence for the Lyapunov exponent for mapping (\ref{B_eq4}) for the control parameter $\epsilon=0.05$ and $p=2$. 

Two important properties of the Lyapunov exponents are worth emphasizing:
\begin{enumerate}
\item{For Hamiltonian systems, $\lambda_1+\lambda_2=0$, or more generally $\sum_{i=1}^N\lambda_i=0$, where $N$ denotes the dimension of the mapping.}
\item{For dissipative systems, the Lyapunov exponents satisfy $\sum_{i=1}^N\lambda_i<0$.}
\end{enumerate}

\end{document}